\documentclass[useAMS,usenatbib]{mn2e}  %[arps,floats,prd,nofootinbib]{revtex4}

\usepackage{dsfont}
\usepackage{amsmath}
\usepackage{amssymb}
\usepackage{graphicx}
\usepackage{verbatim}
\usepackage{natbib}
\usepackage{eucal}
\usepackage{calligra}

\usepackage{amsfonts}
\usepackage{color}
\usepackage[normalem]{ulem}
\usepackage[T1]{fontenc}

\usepackage{times}
\usepackage{epsfig}
\usepackage[usenames,dvipsnames]{xcolor}
\usepackage{url}
\usepackage{caption}
\DeclareMathAlphabet{\mathscr}{OT1}{pzc}{m}{it}

\newcommand{\be}{\begin{equation}}
\newcommand{\ee}{\end{equation}}
\newcommand{\bes}{\begin{equation*}}
\newcommand{\ees}{\end{equation*}}
\newcommand{\bea}{\begin{eqnarray}}
\newcommand{\eea}{\end{eqnarray}}
\newcommand{\beas}{\begin{eqnarray*}}
\newcommand{\eeas}{\end{eqnarray*}}

\newcommand{\Mpc}{\,h^{-1}{\rm Mpc}}

\def \bnabla {{\boldsymbol{\nabla}}}
\def \br {{\bf r}}
\def \bg {{\bf g}}
\def \bv {{\bf v}}
\def \bx {{\bf x}}
\def \bxhat {{\hat \bx}}

\def\apjl{{ApJ}}                % Astrophysical Journal, Letters
\def\apjs{{ApJS}}               % Astrophysical Journal, Supplement
\def\ao{{Appl.~Opt.}}           % Applied Optics
\def\aap{{A\&A}}                % Astronomy and Astrophysics
\begin{document}
\title[Gravitational redshift]
  {Gravitational redshift and asymmetric redshift-space distortions for stacked clusters}
\author[Cai et al.]
{Yan-Chuan Cai$^{1,2}$\thanks{E-Mail: cai@roe.ac.uk}, Nick Kaiser$^{3}$, Shaun Cole$^{2}$ and Carlos Frenk$^{2}$ \\
$^{1}$ Institute for Astronomy, University of Edinburgh, Royal Observatory, Blackford Hill, Edinburgh, EH9 3HJ , UK \\
$^{2}$ Institute for Computational Cosmology, Department of Physics, Durham University, South Road, Durham DH1 3LE, UK \\
$^{3}$Institute for Astronomy, University of Hawaii, 2680 Woodlawn Drive, Honolulu, HI 96822-1839, USA
}
\maketitle
\begin{abstract} 
We derive the expression for the observed redshift in the weak field limit in the observer's past light cone, 
including all relativistic terms up to second order in velocity. We then apply it to compute the cluster-galaxy 
cross-correlation functions (CGCF) using N-body simulations. The CGCF is asymmetric along the 
line of sight (LOS) owing to the presence of the small second order terms such as the gravitational redshift (GRedshift).
We identify two systematics in the modelling of the GRedshift signal in stacked clusters. 
First, it is affected by the morphology of dark matter haloes and the large-scale cosmic-web. 
The non-spherical distribution of galaxies around the central halo and the presence 
of neighbouring clusters systematically reduce the GRedshift signal. 
This bias is approximately 20\% for $M_{\rm min}\simeq 10^{14} {\rm M_{\odot}}/h$, 
and is more than $50\%$ for haloes with $M_{\rm min}\simeq 2\times 10^{13} {\rm M_{\odot}}/h$ at $r>$4 {\rm Mpc}/$h$. 
Second, the best-fit gravitational redshift profiles as well as the profiles of all other 
relativistic terms are found to be significantly different in velocity space compared 
to their real space versions. We discuss some subtleties relating to these effects in velocity space. We also find that the S/N of the GRedshift signal increases with decreasing halo mass. 
\end{abstract}

\begin{keywords}
gravitation -- methods: analytical -- methods: numerical -- large-scale structure of Universe -- galaxies: clusters: general --
cosmology 
\end{keywords}

%%%%%%%%%%%%%%%%%%%%%%%%%%%%
\section{Introduction} 
In general relativity, photons receive a gravitational redshift when climbing out of potential wells. In the weak field limit, the magnitude of the redshift is proportional to the depth of the Newtonian potential $\Phi$.  
Photons from central galaxies sitting at the bottom of the potential well of galaxy clusters are 
expected to be gravitationally redshifted by a larger amount than satellites and other neighbouring galaxies. 
The difference of the gravitational redshift (GRedshift) signal with 
respect to the cluster centre is of the order of 10~km/s. It can in principle be detected by stacking a 
large sample of clusters. This has been predicted by \citep{Nottale1990, Cappi1995, Kim2004} and 
the first few tentative measurements from stacked clusters from SDSS data sets have been reported 
\citep{Wojtak2011, Sadeh2015, Jimeno2015}. 

In observations, the GRedshift signal extracted from stacked clusters is related to the distortion of the 
cluster-galaxy cross-correlation function (CGCF), or $\xi_{\rm cg}$, which originates from the distortions of the observed 
redshifts of galaxies with respect to the cluster centre (which may be the centroid of the
galaxies or may be taken to be the brightest cluster galaxy (BCG)). In theory, ignoring the evolution 
of cosmic potentials and observational systematics, the observed redshift consists of five 
components: (1) the cosmological redshift (2) the 1st order Doppler redshift from the peculiar velocity of the galaxy
(3) 2nd order special relativistic corrections from the peculiar velocity (4) the peculiar gravitational redshift
(5) effects associated with the fact that we observe galaxies on our past light cone. 
The effects of (1) \& (2) result in an observed CGCF that should be front-back symmetric,  
while asymmetry of the CGCF along the line of sight will arise due to the presence of (3), (4) \& (5).
The main goal of this study is to explore these effects on the CGCF
and disentangle the GRedshift effect from them.
There is also an additional effect, (6), the peculiar velocity of galaxies affects their
surface brightness via beaming.  Coupled to any surface brightness dependent selection (such as an apparent magnitude limit) this
results in a bias of the redshift distribution of the selected galaxies at the same order of magnitude.
This last effect, unlike the others, is highly dependent on details of the luminosity
function of the galaxies and how they are selected in the surveys.  Here we shall focus only on those
effects that are independent of how galaxies are selected.

On large scales, the relativistic corrections to the galaxy correlation function and 
the resulting asymmetry of the cross-correlation function between two different `tracer' populations,
in our case clusters and galaxies, 
has been studied in \citet{Yoo2009, McDonald2009, Challinor2011, Bonvin2011,Yoo2012, Croft2013}, 
and in \citet{Bonvin2014} where some other effects such as density evolution and lensing are 
included. Our study will focus on the CGCF at around the scale of clusters and up to tens of Mpc/$h$. 
This is the (quasi-) non-linear regime where some of the theoretical predictions based on 
perturbative methods will break down. It is therefore necessary to employ N-body simulations for this study. 
 
A robust detection of the GRedshift signal may provide a constraint on theories of gravity. 
This requires an accurate prediction of the observed redshift. 
\citet{Wojtak2011}, for example, have modelled the effect by assuming a power-law mass function
for clusters which are individually spherically symmetric and have a NFW \citep{NFW} profile,
and that the observed redshifts are given as the sum of the first order
Doppler shift and the gravitational redshift with respect to the cluster centre.
It was subsequently realised that
several additional physical processes, such as the transverse Doppler redshift, the
past light cone effect and relativistic beaming, would cause additional 
contributions which are generally of the same order of magnitude as the GRedshift signal \citep{Zhao2013, Kaiser2013}
and which complicate the analysis. 
These analyses, however, do not necessarily capture all of the relevant effects that
need to be considered in order to make an accurate prediction. 
One shortcoming is that these analyses are not adequate to treat the
`quasi-linear' regime -- outside the virial radius -- which is observationally
relevant here.  Another is that, of necessity, the quantity that is measured is
a {\em galaxy weighted\/} measurement of the redshift; i.e.\ the mean of the
gravitational redshift, plus other contributions, for galaxies at a given projected
distance from the cluster galaxy centre.  I.e.\ it is not the simple 2-point cluster density-potential
cross correlation function, rather it is a third-order statistic $\langle n_c(0) n_g(\br) \Phi(\br) \rangle / \langle n_c(0) n_g(\br) \rangle$, where $n_c(0)$, $n_g(\br)$ and $\Phi(\br)$ are the number density of central galaxies at the origin, the number density of galaxies at $\br$ and the peculiar Newtonian potential at $\br$.
Here we use N-body simulations to attempt to remedy these shortcomings.

The outline of the paper is as follows:  
In the next section, we derive an expression for the observed redshift accurate to second order
in the velocities (Hubble and peculiar) and to first order in the peculiar potential
and allowing for the fact that we observe galaxies on the past light cone. This
provides the redshift in terms of quantities defined on a hyper-surface of
constant time, which is useful as the simulations provide snap-shots of
the galaxy positions, velocities and the peculiar gravity on such hyper-surfaces.

We analyse the simulations in \S\ref{sec:stacks}.  This analysis reveals and quantifies
two important new complicating factors.  The first has to do with the fact
that while, in a composite sense, clusters are spherically symmetric,
individual clusters are aspherical and their surroundings are highly aspherical
owing to the presence of neighbouring clusters.  Coupled with the fact
that the quantity one most naturally measures is the {\em galaxy weighted\/}
redshift and clumps of galaxies are correlated with potential wells
this results in a systematic bias which causes the weighted potential to 
increase more slowly with distance from the cluster centre than one would
expect from simple models invoking an ensemble of spherical NFW profile clusters.
The second effect has to do with the fact that the galaxies are observed
in velocity space rather than in real space. 

\section{Relative redshifts on the past light cone}

We summarise the source of distortions to the observed redshifts below.

To the lowest order in peculiar velocity and potential, the distortion is associated only with the Doppler 
redshifts from the line-of-sight component of the peculiar velocities. The redshift of a galaxy is
\begin{equation}
cz=Hx+v_x,
\end{equation}
where $x$ is the cosmological comoving, or conformal, distance, $v_x$ is the line-of-sight peculiar velocity, $H$ is the Hubble constant and $c$ is the speed of light.

In General Relativity, gravitational redshift will add to the observed redshift by the amount that is proportional to the depth of the Newtonian potential $\Phi$. The gravitational redshift is of the order of 10 km/s for galaxy clusters with mass $M\sim10^{14}{\rm M_{\odot}}/h$. In the context of special relativity, \citet{Zhao2013} realised that the transverse Doppler redshift term, dependent on $v^2$, should also be added. It is guaranteed to be of the same order as the gravitational redshift term. \citet{Kaiser2013} showed that there is an additional effect that is of order $v^2$ that comes about because the galaxies are
observed on the past light cone.

In this section we establish the connection between the Hubble and peculiar velocities
of galaxies (or particles in an N-body simulation) in the vicinity of a cluster and
the redshift, as would be measured by some distant observer, in the first instance, relative to the redshift of 
of a stationary reference source that lies at the origin of coordinate system.  We then generalise
this to give the redshift relative to the cluster centre.

Since these relative redshifts are very small we may analyse this
using Newtonian gravity with gravitationally induced wavelength shifts $\delta \lambda / \lambda_{\rm em} = - \Phi / c^2$,
where $\delta \lambda \equiv \lambda_{\rm obs} - \lambda_{\rm em}$,
and using special relativity to compute the Doppler shifts.  
Since  these wavelength shifts are multiplicative we can simply deal with this as treating  their logarithms as additive.
Furthermore, since the (total) potential 
is of the order of the square of the total velocity (i.e. Hubble plus peculiar),
it is sufficient here to work to second order in velocities and first order
in the potential.  Also, since the potential will generally be evolving on
the dynamical time-scale and the velocities are highly non-relativistic one 
may ignore the evolution of the potential in the relatively tiny light
travel time.

More specifically the relative redshift may be calculated in terms of peculiar
velocities and the peculiar potential -- that is the solution of Poisson's
equation with the density perturbation as the source term and which is what appears
in the equations of motion that are solved in N-body simulations -- as follows:
First we may calculate the relative redshifts that would be observed in the
fictitious situation where the density of the universe is unperturbed but
where one is observing a set of particles that have peculiar velocities.
This is given by (one plus) the cosmological redshift -- which is just inversely proportional
to the scale factor at the time of emission -- multiplied by the relativistic
Doppler shift for a moving particle with respect to a co-moving particle (i.e.\ one
with vanishing peculiar velocity).  Note that the peculiar velocity here
is the peculiar velocity at the emission time, whereas what we are supplied
with most conveniently is the output of N-body simulations on a hyper-surface
of constant proper time. So it is necessary to allow for the `Hubble drag'
which causes a change of peculiar velocity with time. If we now `switch on'
the effect of gravity we need to include the lowest order gravitational
redshift by adding the appropriate Newtonian peculiar potential to the
fractional frequency shifts and we need to allow for the fact that there
is not just Hubble drag but also peculiar gravitational acceleration which changes the peculiar velocity.

Let us suppose we are given a set of galaxies coordinates and velocities (or those of
the particles in a N-body simulation) on a constant proper-time hyper-surface. 
More specifically let us assume that for each galaxy we have 
the position $\br$, this being the co-moving coordinate times the scale factor $a$; the peculiar velocity $\bv = d \br / dt$ (from which
we can obtain the conformal velocity $\dot \br \equiv d \br / d \eta = a \bv$), where conformal time
is defined, up to a constant, by $d \eta = dt / a(t)$.  

Let us also assume
that we are provided with the peculiar
potential $\Phi$ and its gradient $\bg = - \bnabla_\br \Phi$, again at some given conformal time $\eta = \eta_0$.

We will use units such that $c = 1$ temporally, and put back $c$ for the final expression of our derivation.
If we set $a = 1$ at the output time then $\bv$ and $\dot \br$ are identical at that time
and separations in $\br$ are proper separations in physical units.

\begin{figure}
\scalebox{0.6}{
\includegraphics[angle=0]{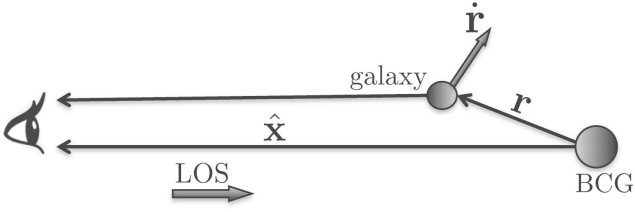}}
\caption{Illustration of the BCG-galaxy system. The observer is on the left, assumed to be static in conformal coordinates with respect to the BCG.  
${\bf r}$ is the conformal position of the galaxy with respect to the cluster centre, and ${\bf \dot r }$ is its conformal velocity. 
Photons received by the observer at the conformal time $\eta_0$ are emitted at a different conformal time from the galaxy and the BCG.
During this interval of look-back time, the Universe expands, the galaxy moves and may be accelerated with respect to the BCG. 
These give rise to the second order terms in Eq. ~(\ref{Eq:full2}).}
\label{Fig:Illustration}
\end{figure}

Extending off the time-slice $\eta = \eta_0$, a galaxy will have trajectory
\begin{equation}
\br(\eta) = \br + (\eta - \eta_0) \dot \br + \ldots
\label{eq:worldline}
\end{equation}
where $\br$, and $\dot \br$ without an argument indicate the values at $\eta_0$
and $\ldots$ indicates terms that are of second or higher order 
in conformal look-back time $\Delta \eta \equiv \eta - \eta_0$.

As illustrated in Fig.~\ref{Fig:Illustration}, we will place the observer event at some large distance along the (minus) $x$ axis, 
and at the time such that the observer receives photons that left the origin (which we
will ultimately take to be the centre of the cluster) at time $\eta_0$.
The equation of the surface in $\br$-space that contains the points on the observer's past light cone
with conformal time $\eta$ is
\begin{equation}
\eta = \eta_0 - \bxhat \cdot \br(\eta) + \ldots
\label{eq:lightcone}
\end{equation}
where $\bxhat$ is the unit vector parallel to the $x$-axis and
where we are ignoring the fact that the coordinate speed of light is not exactly unity
because of the metric perturbations (this introduces errors of order $v \times \Phi$ which we may safely neglect). 
This formula gives the conformal emission time of a photon from a particle at relative
position $\br$ that is received at the same time as a photon which leaves the origin
at time $\eta_0$.

For simplicity here we are making the `plane-parallel' approximation, which is
valid for sufficiently distant clusters. 

Substituting (\ref{eq:worldline}) in (\ref{eq:lightcone}) yields the conformal look-back time
in terms of $\br$ and $\dot \br$:
\begin{equation}
\Delta \eta = - \bxhat \cdot \br / (1 + \bxhat \cdot \dot \br) = - \bxhat \cdot \br 
+ (\bxhat \cdot \br) (\bxhat \cdot \dot \br) + \ldots
\end{equation}
or, with $x = \bxhat \cdot \br$ and $\dot x = \bxhat \cdot \dot \br=v_x$
\begin{equation}
\Delta \eta = - x + x \dot x + \ldots
\label{eq:Deltaeta}
\end{equation} 

We may use this to calculate the (inverse) redshift associated with
the expansion of the universe during this look-back interval:
\begin{equation}
(1 + z)^{-1} = \frac{a(\eta)}{a(\eta_0)} = 1 + \frac{\dot a}{a} \Delta \eta + \frac{1}{2} \frac{\ddot a}{a} (\Delta \eta)^2 + \ldots
\end{equation}
or
\begin{equation}
1 + z = 1 - \frac{\dot a}{a} \Delta \eta + \left[\left(\frac{\dot a}{a}\right)^2 - \frac{1}{2} \frac{\ddot a}{a} \right] (\Delta \eta)^2 + \ldots
\end{equation}
or with $\Delta \eta$ given by (\ref{eq:Deltaeta})
\begin{equation}
1 + z = 1 + \frac{\dot a}{a} x - \frac{\dot a}{a} x \dot x 
+ \left[\left(\frac{\dot a}{a}\right)^2 - \frac{1}{2} \frac{\ddot a}{a} \right] x^2 + \ldots
\label{eq:oneplusz}
\end{equation}

This is {\em not\/} the redshift of a real galaxy with time-slice
position $\br$ and velocity $\dot \br$
(at the time $\eta$ when it intercepts the observer's past light cone), rather it is the redshift of
a stationary source that is co-located with that galaxy at that time 
relative to a stationary source at the origin $\br = 0$ in a fictitious
universe with no structure and therefore no peculiar gravitational redshift. 
To obtain the the redshift of the actual
particle of interest we need to multiply (\ref{eq:oneplusz})
by the appropriate Lorentz boost factor and we need to include the peculiar gravitational redshift.

The Doppler shift (the redshift of the emitting galaxy as seen by a co-located stationary observer) is \citep{Einstein1907}
\begin{equation}
(1 + z)_{\rm Doppler} = \frac{1 + \dot x}{\sqrt{1 - v^2}} = 1 + \dot x + v^2 / 2 + \ldots, 
\label{eq:Dopplerterm}
\end{equation}
but here $\dot x$ is the peculiar velocity at the time of emission, which differs (at 2nd order) from
the velocity at the output time $\eta_0$.  The equation of motion for the peculiar velocity is
\begin{equation}
\dot \bv = \bg - H \bv
\end{equation}
where $\bg$ is the peculiar acceleration and the second term is the `Hubble drag' term that arises
because peculiar velocities are defined to be with respect to the expanding (constant co-moving coordinate) observers.
Thus the line-of-sight velocity appearing in (\ref{eq:Dopplerterm}) is
\begin{equation}
\dot x(\eta) = \dot x(\eta_0) - (g_x - H \dot x) x
\end{equation}
where we have used $\Delta t = \Delta \eta = -x$.

Multiplying (\ref{eq:oneplusz}) and (\ref{eq:Dopplerterm}) and keeping up to 2nd order terms and
adding the peculiar gravitational redshift gives, for the redshift of the galaxy with
respect to that for a stationary emitter at the origin,
\begin{equation}
\begin{aligned}
\label{Eq:full}
cz= & Hx+v_x+v^2/2c-\Phi/c \\ 
& -xg_x+Hxv_x/c+\left[H^2-\ddot a/(2a^2)\right]x^2/c,
\end{aligned}
\end{equation}
where we have put back the speed of light. 
The above equation fully accounts for the observed redshift relative to a stationary emitter on the past
light cone to second order (if the potentials are not evolving). 
We call the total distortion to the Hubble term induced by all the other terms the ultimate redshift-space distortion (uRSD). 

The above formula gives the redshift of a galaxy (or particle in a simulation) relative to a stationary source
lying at $\br = 0$. More observationally relevant is the redshift relative to the centre of the cluster.
This might be defined to be the brightest cluster galaxy (BCG), or it may be defined to be the centroid
of the cluster members.  The above formula can be used to obtain the redshift of the BCG, and one might naively imagine
that the relative redshift of the galaxy relative to the BCG would be the difference of these.  But this is
not the case; at least when working to 2nd order precision.  The relevant relative redshift is 
$1 + \delta z = \lambda_{\rm obs} / \lambda'_{\rm obs} = (1 + z) / (1 + z')$ where $ \lambda'_{\rm obs}$
is the observed wavelength for light received from the centre and $z'$ is the corresponding redshift.
Because $z'$ appears in the denominator, we cannot simply take $\delta z = z - z'$. 

In order to deal with this situation, or the yet more complicated situation where the centre of the
cluster is the centroid of the members it is more convenient to work in terms of $z_*$, the logarithm
of $1 + z$.  As the relative $z_*$ between the galaxy is just the difference of $z_*$ for these objects
relative to the reference source.  Also, the $z_*$ of the centroid is just the average of the $z_*$
values for the cluster members.  At second order, 
\begin{equation}
\label{eq13}
z_* = z - ((Hx + v_x)/c)^2 / 2, 
\end{equation}
and we have
\begin{equation}
\begin{aligned}
\label{Eq:full-star}
cz_*= & Hx+v_x+v^2/2c - v_x^2 / 2c -\Phi/c \\ 
& -xg_x+\left[H^2-\ddot a/(a^2)\right]x^2/(2c),
\end{aligned}
\end{equation}
To estimate the impact of those second order terms for real observations, there is no unique way, as it depends on what convention 
the `observed redshift' is adopted. For example, \citet{Wojtak2011} took the LOS `velocity' of a galaxy {\it wrt} the BCG as 
${\Delta V_{LOS}} = c\frac{z-z_{\rm c}}{1+z_{\rm c}}$. In terms of $z_*$, it becomes
\begin{equation}
\label{Wt}
\frac{\Delta V_{LOS}}{c}  = \frac{e^{z_*}-e^{z_{c*}}}{e^{z_{c*}}} \approx \Delta z_* + \frac{1}{2} \Delta z_*^2, 
\end{equation}
where $\Delta z_*=z_*-z_{*c}$ and the subscript $c$ denotes quantities for the BCG. However, if one uses $(z-z_{\rm c}) / (1 + z)$ instead of having 
$(1 + z_c)$ in the denominator, then Eq.~(\ref{Wt}) would become 
\begin{equation}
\frac{\Delta V_{LOS}}{c}  =1 -\exp(-\Delta z_*)\approx \Delta z_* - \frac{1}{2} \Delta z_*^2, 
\end{equation}
which is different from Eq.~(\ref{Wt}) at the second order. 
Nevertheless, in this work, we choose the convention of  Eq.~(\ref{Wt}) as an example for illustration.
Combining Eqs~(\ref{eq13}-\ref{Wt}), we have
\begin{equation}
\begin{aligned}
\label{Eq:full2}
\Delta v_{LOS}= & H x + \Delta v_x+ \Delta v^2 /(2c)+[(\Delta v_x)^2 - \Delta v_{x}^2]/(2c)- \\
& \Delta \Phi /c - xg_x + Hx\Delta v_x/c+ (H^2-\ddot{a}/a^2/2)x^2 /c .
\end{aligned}
\end{equation}
$\Delta x$ and $\Delta v_x$ are differences of a galaxy's LOS distance and peculiar velocity {\it wrt} the BCG respectively, so $\Delta x=x$ and $\Delta v_x \not= v_x$ by definition. $\Delta v_x=v_x-v_{xc}$,  $\Delta v^2=v^2-v^2_c$ and $\Delta \Phi =\Phi- \Phi_c$.

The various terms in the uRSD, Eq.~(\ref{Eq:full}), can be understood as follows.
\begin{itemize}
\item The first two terms on the RHS are the Doppler shift from the total (i.e.\ Hubble + peculiar) velocity.
\item We then have the transverse Doppler effect and the peculiar gravitational redshift.
\item Next we have minus the product of the line-of-sight displacement and the line-of-sight acceleration; 
these tend to be anti-correlated for over-dense systems and combine to give the (positive redshift) effect shown in 
\citep{Kaiser2013}, but in Section~\ref{sec:2nd_Order_Term}
we will see the situation is more complicated in velocity space.

\item Next we have a second order term $H x v_x/c$ that is the product
of the Hubble and peculiar velocities. In the virialised region these will be uncorrelated,
but in the outskirts of a cluster they will be anti-correlated so  should give a negative contribution
to the mean redshift. Again, the situation in velocity space and further from the cluster centre 
may be different.
\item We then have the quadratic term (in $x$) that comes from the the combination of the
background gravitational redshift and Doppler effects (it is present even if $\bv$ and $\Phi$ are
zero).  In a situation where the density of galaxies is constant in real space, this will
introduce, at leading order, a linear ramp in the density.  However, in analyses of 
gravitational redshift such as those of \citet{Wojtak2011} and \citet{Jimeno2015} this gets removed because they fit for the local large-scale gradient using the
density of galaxies well separated in velocity from the cluster.  Similar effects arise from
the fact that the cluster will be at finite distance, so a beam through the cluster in which
the distribution of galaxies is measured will
be broadening, and also because of variation of the selection function.  We will assume that
the process for fitting the background density ramp has removed all of these.
\end{itemize}
%The background potentials are neglected as they are small (sub-km/s) at the scale of our interests. 

As mentioned, there is one final complication in that the surface brightness of a galaxy
at a given distance and light emission time depends on the peculiar velocity.  This couples
to the selection criterion. One could deal with this using the `Poisson sample' model in which we
assume that galaxies in a given volume element are drawn from the luminosity function,
and are then selected according to observational criteria, and where the overall normalisation
includes the space density of haloes as a multiplicative factor.  The `Doppler boosting' modulation can be incorporated
by giving a weight to the haloes extracted from the simulations.  We note that, unlike the
other effects, this is sensitive to exactly how galaxies are selected, which in turn is a function
of distance to the cluster.  

On the RHS of the above equation, apart from the first two terms, which give rise to the conventional RSD, all the other terms will cause asymmetry in the CGCF. The RSD signal coming from the $\Delta v_x$ term is expected to be dominant over all other effects. 
Recovering the asymmetry signal from the observed CGCF resulting from the uRSD is nontrivial.
Our goal is to use N-body simulations to 
quantify each of these terms and so determine the contamination of the
GRedshift signal.
Before we quantify the redshift space distortion of the CGCF
caused by each
effects of the of Eq.~(\ref{Eq:full2}) 
we first study the GRedshift signal using the full 3D, real space
information of particles in our N-body simulations. We 
show that even in this ideal situation, there is a subtle systematic effect when one assumes spherical symmetry when stacking. 

%%%%%%%%%%%%%%%%%%%%%%%%%%%%    
\begin{figure*}
\begin{center}
% fig 1
\vspace{-1.0cm}
\scalebox{0.48}{
\includegraphics[angle=0]{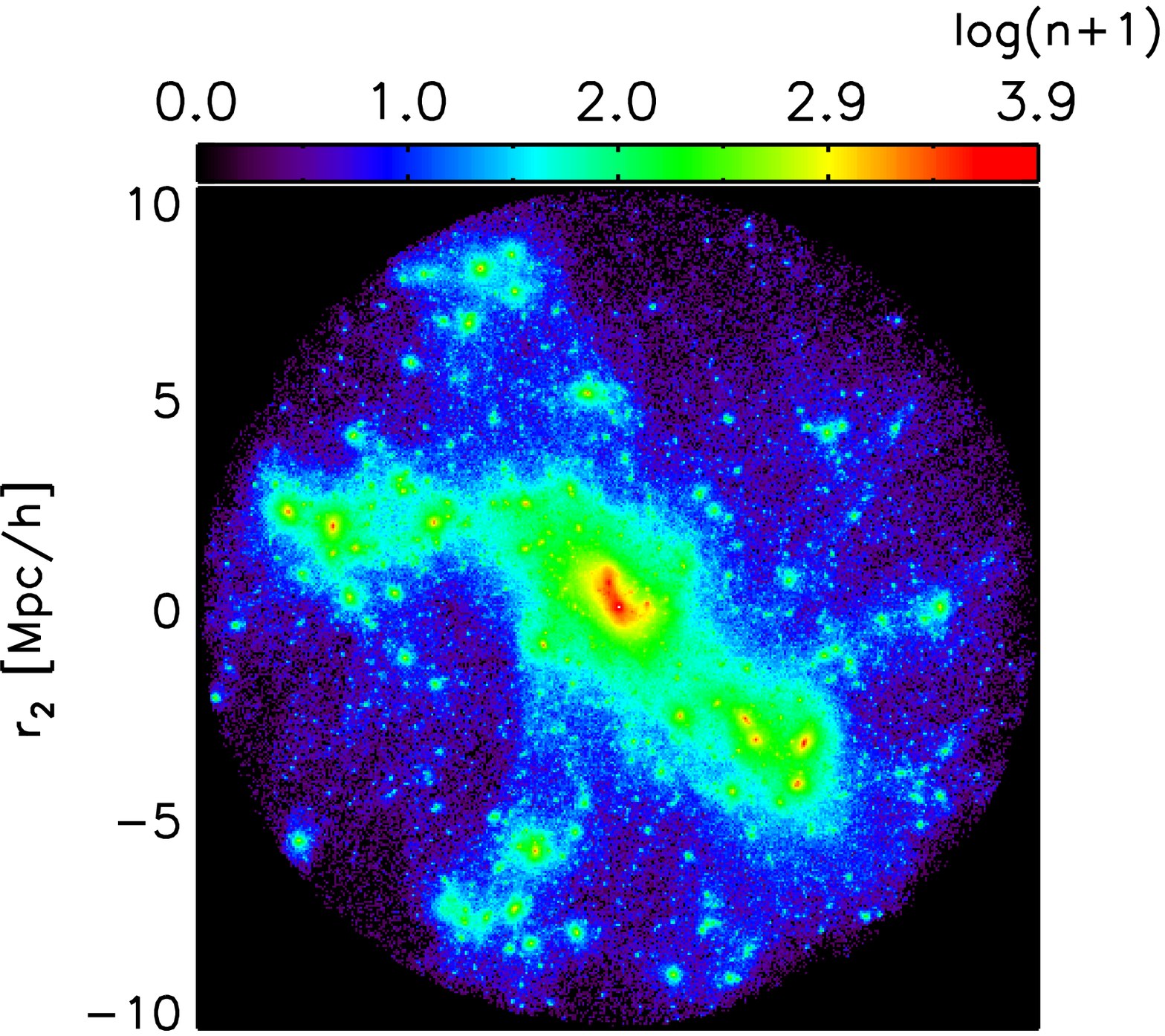}}
\hspace{-3.0 cm}
\vspace{-2.87cm}
% fig 2
\scalebox{0.48}{
\includegraphics[angle=0]{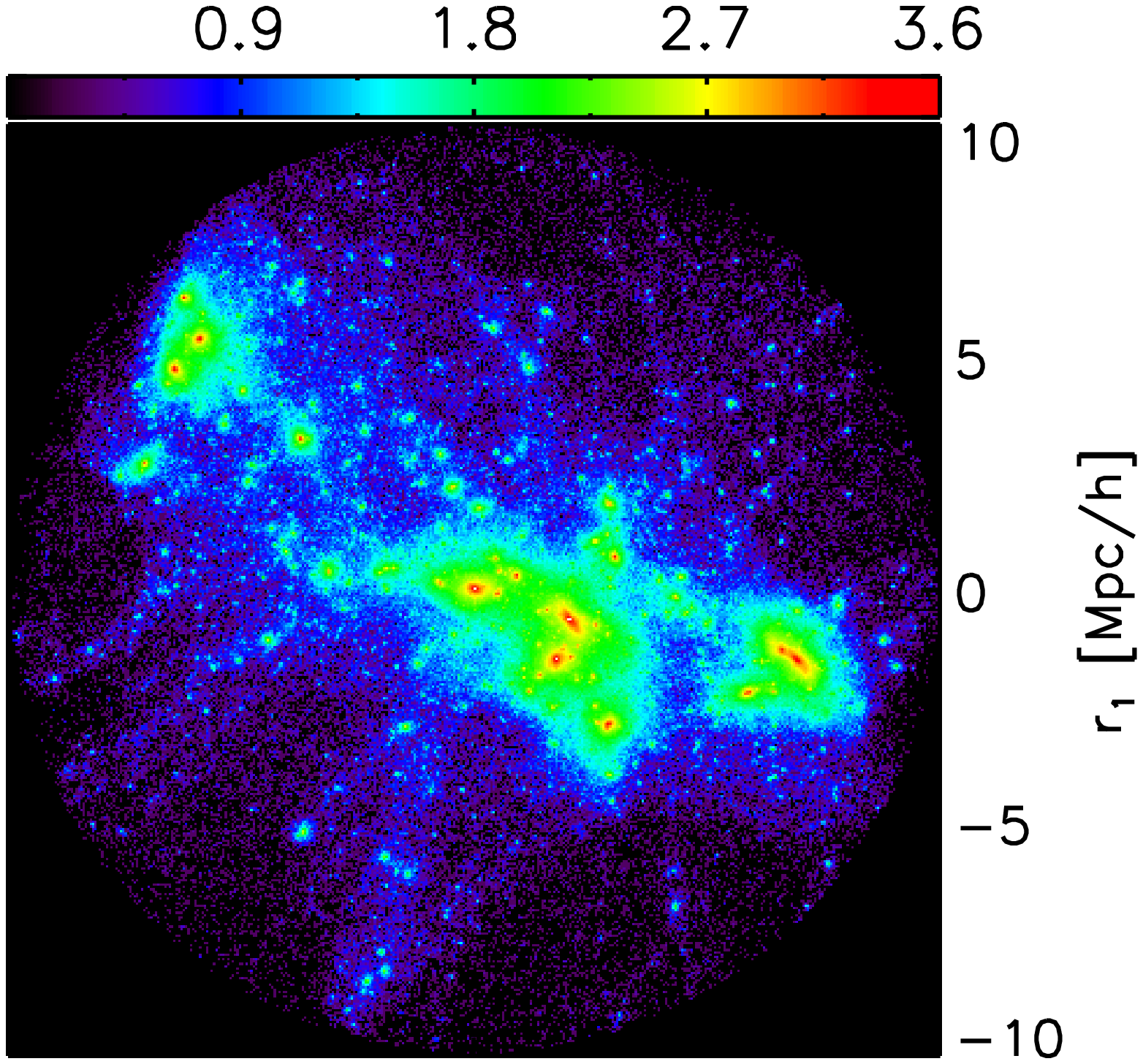}}
\vspace{-2.87cm}
% fig3
\scalebox{0.48}{
\hspace{0.12 cm}
\includegraphics[angle=0]{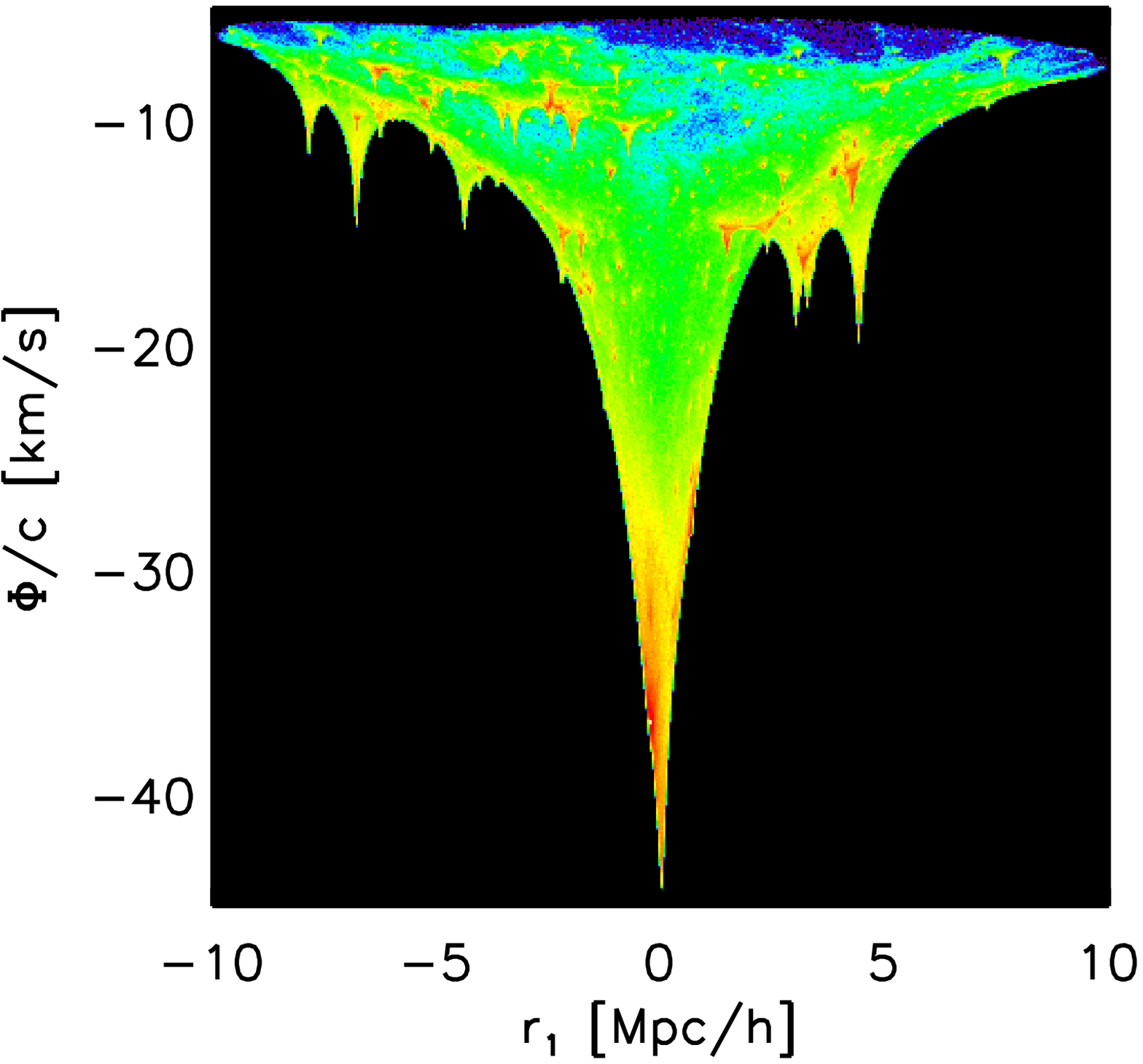}}
\hspace{-3.0 cm}
\vspace{1.4cm}
%fig 4
\scalebox{0.48}{
\includegraphics[angle=0]{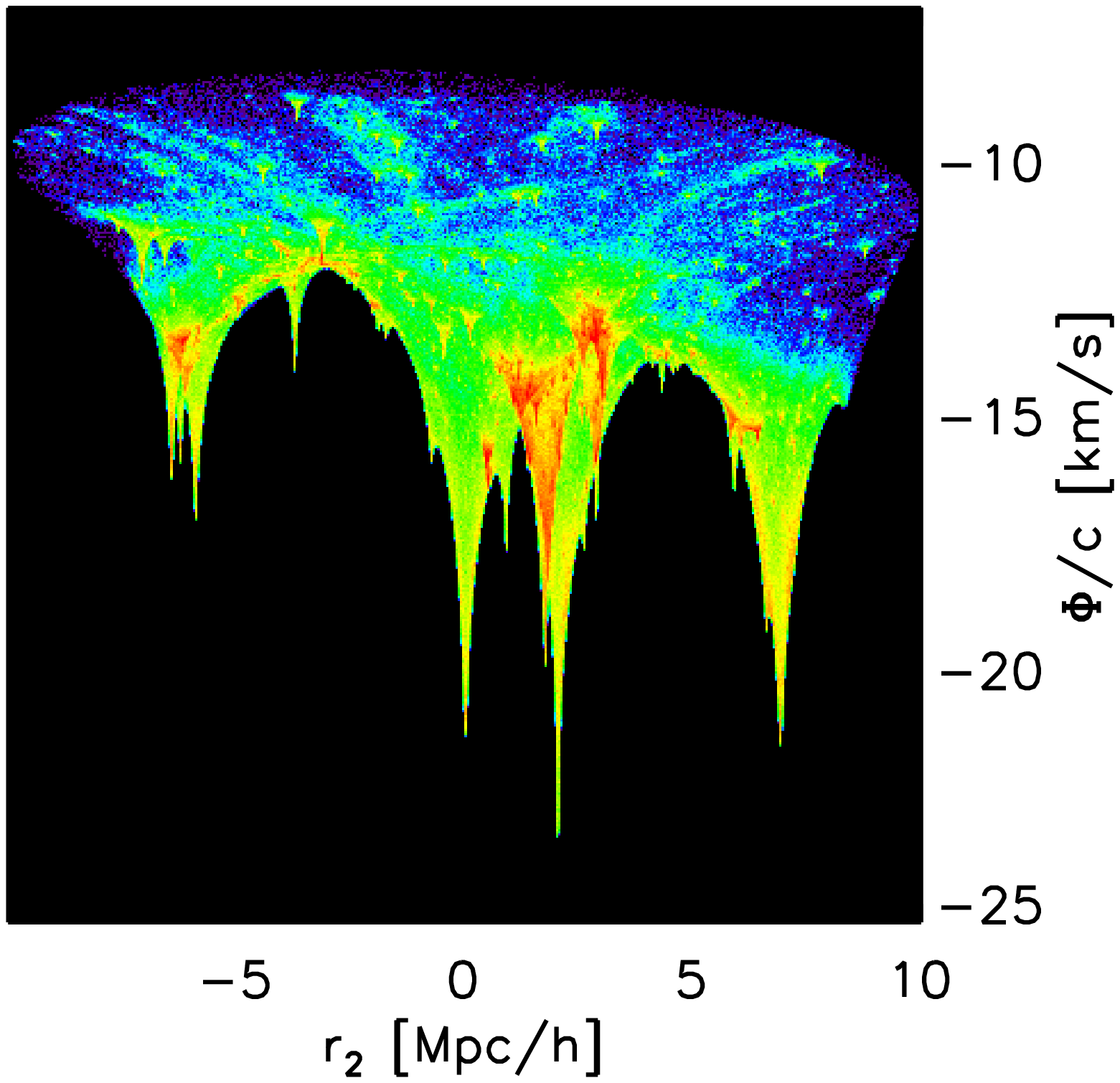}
}
\vspace{1.3cm}

\hspace{-1.6 cm}
\scalebox{0.41}{
\includegraphics[angle=0]{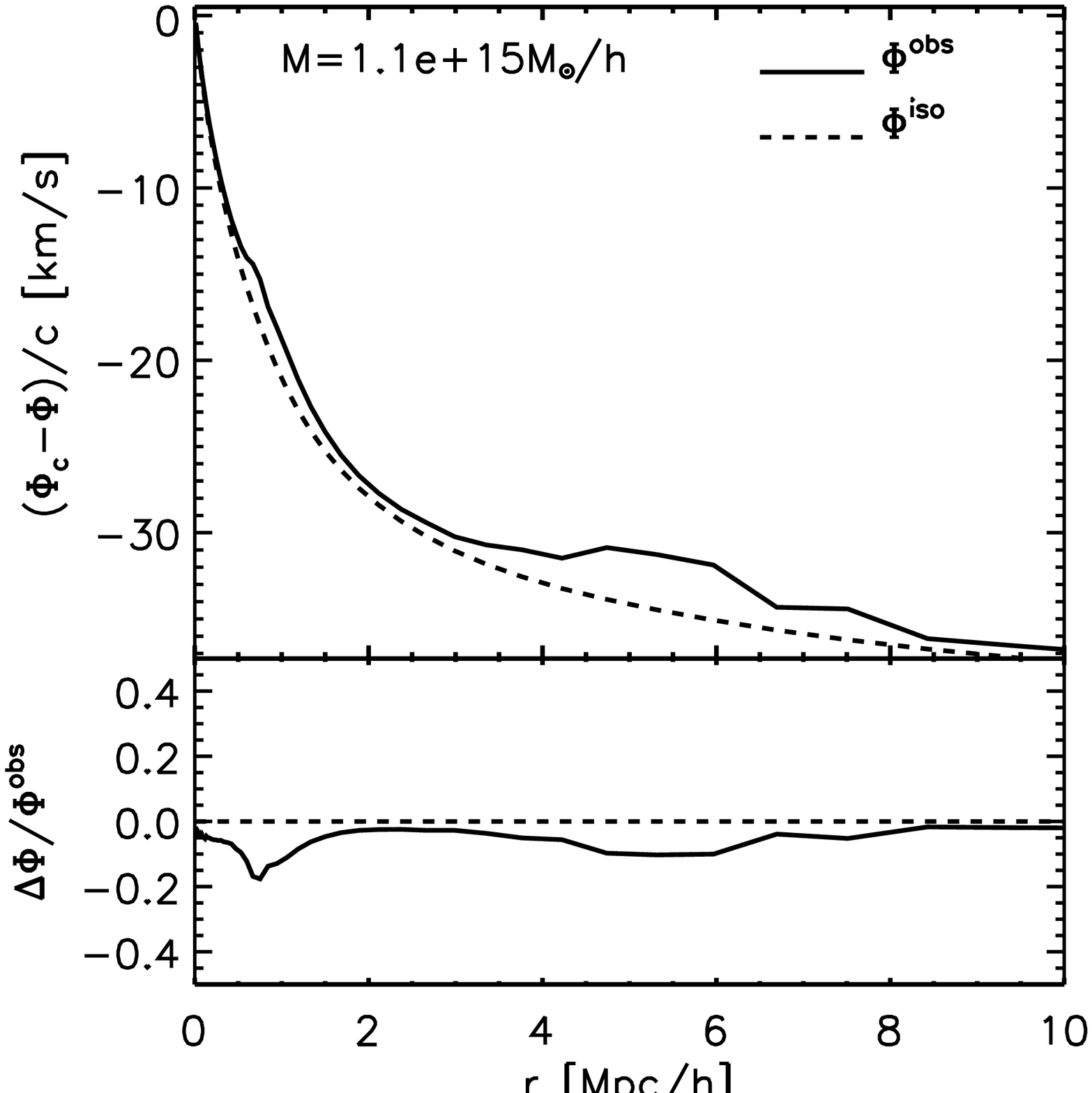}}
\hspace{-1.63 cm}
\scalebox{0.41}{
\includegraphics[angle=0]{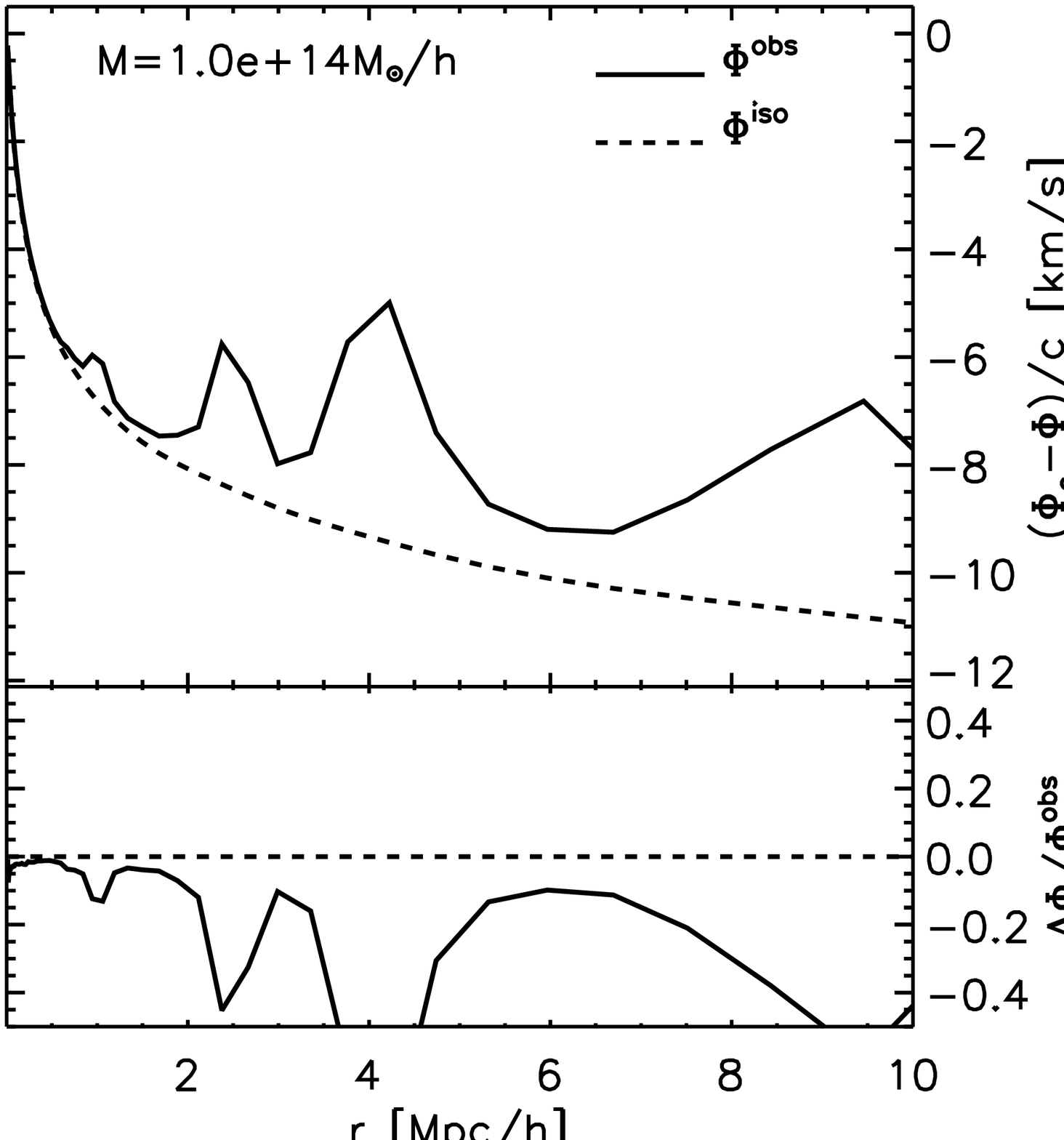}
}
\caption{Top row: particle distributions within a 10 Mpc$/h$ radius of the main halo 
centre projected along one major axis of the simulation box.
The colour displays the number of particles in each pixel, $n$, 
as indicated by the colour bar.
Middle row: The same regions and colour coding as the top panels but
now showing the value of the potential of each particle on the $y$-axis.
Sub-haloes and neighbouring structures generate local potential minima. 
Bottom row: the gravitational redshift profiles relative to the cluster centres. 
The dashed lines shows the spherical averaged profile, $\Phi^{\rm iso}$, which is the same as obtained by
isotropic weighting from the halo centres. Sub-haloes and neighbours cause
the mass weighted profiles $\Phi^{\rm obs}$ to be biased low compared to spherical averaging. 
This is similar to observations where the observed profiles are weighted by galaxies.
}
\label{Fig:Individual}
\end{center}
\end{figure*}

%\section{The origin of biases for cluster potential profiles}
\section{Testing the spherical assumption for stacked clusters}
\label{sec:stacks}

The first measurement of the gravitational redshift signal from
stacked clusters has been conducted by \citet{Wojtak2011} using the SDSS MaxBCG group catalogue from \citep{Hao2010}. 
The idea is that the BCG is likely to live close to the bottom of the potential well of
the host halo, while other galaxies
(satellites and field galaxies) further away from the centre of the
halo tend to occupy locations where the gravitational potential is shallower.
Therefore, there are relative blue shifts of the spectra coming from other galaxies relative to that of the BCGs seen by the observer. 

In observations, the gravitational redshift signal of stacked clusters arises from 
the galaxy number weighted gravitational potential profile,
\begin{equation} 
\label{Eq:obs}
\bar \Phi^{\rm obs}(r)=\frac{\int (dN_c/dM) dM{\int n_{c,g}({\bf r})[\Phi_{\rm c}-\Phi({\bf r})]d\Omega}}{ \int (dN_c/dM)dM\int n_{c,g}({\bf r})
d\Omega},
\end{equation}
where $\Phi_{\rm c}$ and $\Phi({\bf r})$ are the Newtonian
potentials at the centre of the cluster and at the position ${\bf r}$, $dN_c/dM$ is the number density of clusters per unit mass (i.e. the cluster mass function) and $n_{c,g}({\bf r})$
is the number (density) of observed galaxies in the cluster at ${\bf r}$. It
is important to note that ${\bf r}$ is a 3D vector, and not a scalar.
Previous studies in this subject usually take $r$ as a scalar, which 
implicitly assumes spherical symmetry for each cluster as well as 
for the stacked cluster composite. It is reasonable to expect that the stacked cluster will be
close to being spherically symmetric as long as the sample is large. However, individual clusters are not spherically symmetric
and contain substructures. Hence in the stack 
their potentials will be weighted
more strongly at the locations where there are more galaxies (and more mass), and since more mass is 
associated with deeper local potentials this will bias the 
weighted potential relative to the spherically averaged potential.
This is one of the key points
that we aim to address. By writing down
${\bf r}$ as a 3D vector, the above expression gives the mass
weighted (or galaxy number weighted) potential profile.

To be explicit, for the mass weighted case, the averaged potential at
a given $r$ for the stacked clusters is affected by the contribution of
each cluster system in three different ways.
First, the composite is weighted by the number of galaxies contributed by each cluster at each $r$. More massive clusters generally contribute more weight.
Second, within each spherical shell, the potential is weighted more strongly in  the directions where there is more mass and more galaxies.
This subtle effect can bias the average potential. 
Third, the potential profiles are weighted by the number of
clusters. This is represented by the outer integral over the halo mass
function in Eq.~(\ref{Eq:obs}), where in observations, the selection function should also be incorporated properly.

Moreover, when measuring the gravitational redshift profiles at a
few times the virial radius of the main halo, it is over-simplistic to extrapolate an analytic halo profile such as the NFW profile \citep{NFW}  to such large radii.
The presence of neighbouring galaxies and clusters will generate local potential minima. They will alter the shape of the potential profile. Naively, one may expect that these are random fluctuations and they will cancel out when averaging over a large sample of clusters. 
However, the fact that local potential minima
are correlated with the local overdensity of galaxies
will mean the galaxy weighted potential will be biased
by such fluctuations in the potential.

Eq.~(\ref{Eq:obs}) is essentially what one will measure from observations, whereas
for the modelling, spherical symmetry is usually assumed for each
cluster. This is equivalent to dropping the angular dependence from Eq.~(\ref{Eq:obs}):
\begin{equation} 
\label{Eq:iso}
\bar \Phi^{\rm iso}(r)=\frac{\int (dN_c/dM) {n_{c,g}(r)[\Phi_{\rm
c}-\Phi(r)]dM}}{ \int (dN_c/dM) n_{c,g}(r)dM}.
\end{equation}
In this case, the potentials are weighted equally in all directions, but at each 
$r$, we  retain the relative weighting of clusters of different masses.
We will refer to this case as `isotropic with halo mass weighting'.

A even more naive model for the potential profile is to give equal
weight to each spherical shell for each cluster. In this case, Eq.~(\ref{Eq:iso}) 
is becomes:
 \begin{equation} 
 \label{Eq:idl}
\bar \Phi^{\rm idl}(r)=\frac{\int (dN_c/dM) [\Phi_{\rm c}-\Phi(r)]dM}{ \int
(dN_c/dM)dM},
\end{equation}
which we refer to this idealised case as equal weighting. 
We use cosmological N-body
simulations to see how, in practice,
the three stacked potential profiles of Eqs~(\ref{Eq:obs}-\ref{Eq:idl}) differ.

\begin{figure*}
\begin{center}
\advance\leftskip -1.0cm
\scalebox{0.5}{
\includegraphics[angle=0]{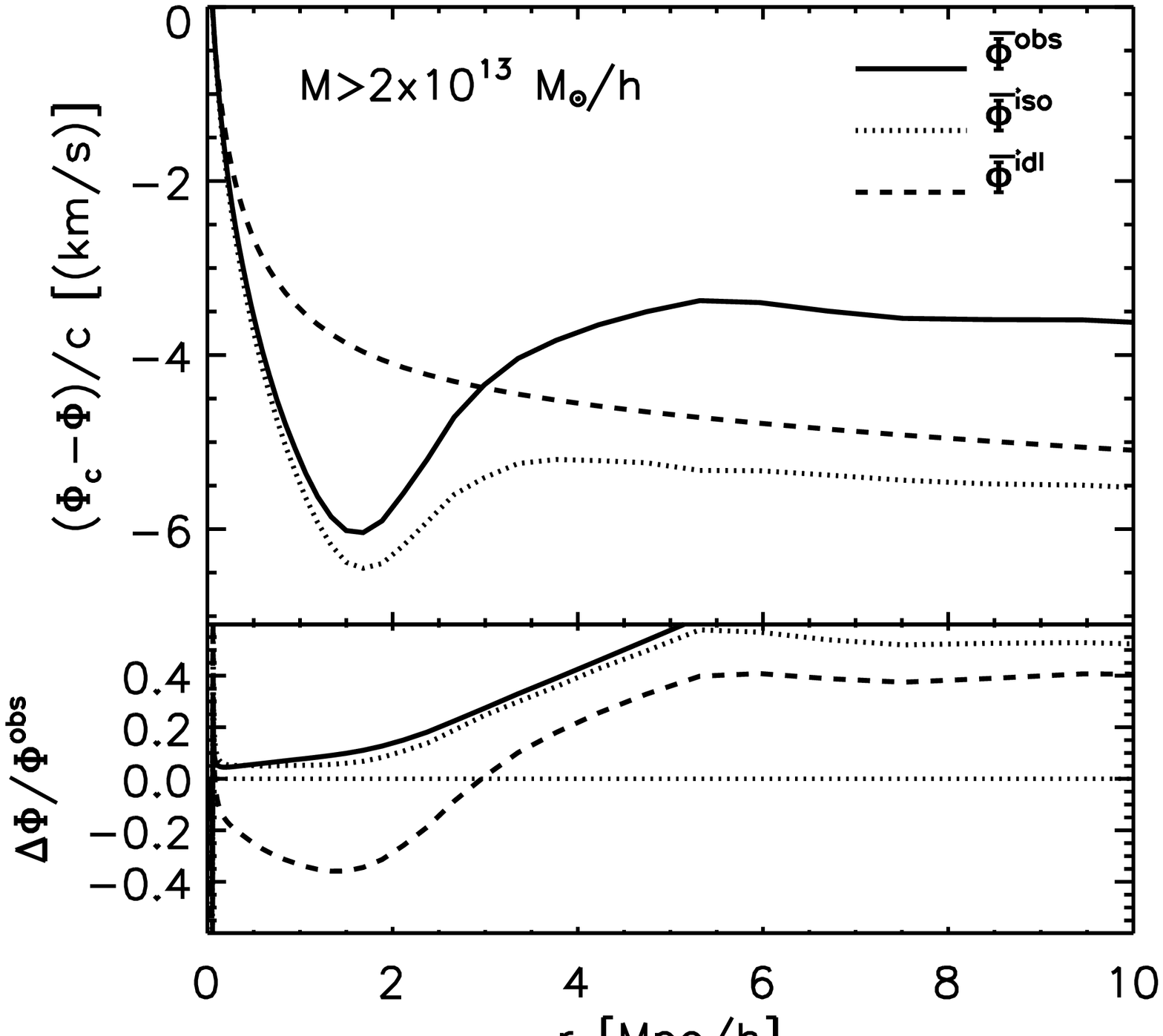}
\includegraphics[angle=0]{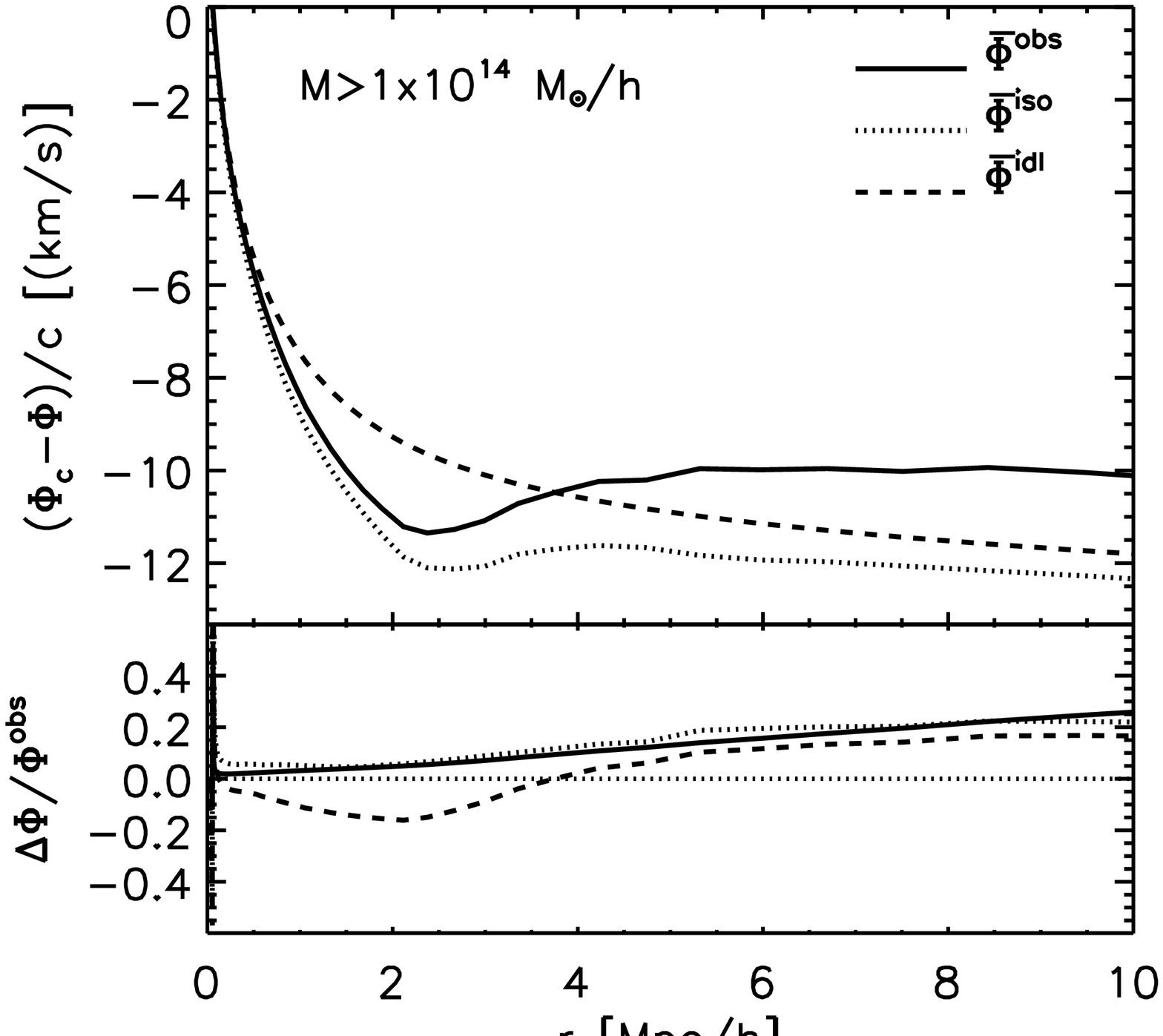}
}
\caption{The gravitational potential profiles for stacked haloes in different
mass bins as indicated in the legends. Potential values for halo centres, $\Phi_{\rm c}$, are approximated by averaging those of all particles
within a core radius of $r_c=3$~kpc/$h$ around the location of each most bound particle.  
Solid lines show results from the mass weighted, $\bar \Phi^{\rm obs}$ of Eq.~(\ref{Eq:obs}). The dotted lines are for the isotropically weighted case, $\bar \Phi^{\rm iso}$ of Eq.~(\ref{Eq:iso}). The dashed lines are for the idealised case of $\bar \Phi^{\rm idl}$ from Eq.~\ref{Eq:idl}, where  we additionally give equal weight to each halo rather than weighting them according to the mass they contribute to the shell. Bottom panels show the fractional differences 
$(\bar \Phi^{\rm obs}-\bar \Phi^{\rm iso})/\bar \Phi^{\rm obs}$ (dotted lines) and  $(\bar \Phi^{\rm obs}-\bar \Phi^{\rm idl})/\bar \Phi^{\rm obs}$ (dashed lines).
The absolute differences between $\bar \Phi^{\rm obs}$ and $\bar \Phi^{\rm iso}$  are found to follow a simple linear function when the radius is rescaled by $r_{200}$, i.e. $\Delta \Phi/c \approx 0.25r/r_{200}$~km/s, as shown by the solid curves in the bottom panels.}
\label{Fig:Stack}
\end{center}
\end{figure*}

\section{The simulation set up}

We use the Millennium simulation for our study \citep{Springel2005}. The simulation was run in the concordance $\Lambda$CDM model,
with $\Omega_{\rm m}=0.25$,  $\Omega_{\rm \Lambda}=0.75$, $h=0.73$,
$\sigma_8=0.9$ and $n=1$.
It has $2160^3$ particles in a box of 500 Mpc/$h$ on a side.
The particle mass is $8.6\times 10^8 {\rm M_{\odot}}/h$. We focus on haloes with the mass of 
$M>10^{13} {\rm M_{\odot}}/h$. They have at least $10^4$ 
particles. The softening length of the force is 5~kpc/$h$. The high 
resolution of the simulation enables us to probe the gravitational potential profiles deep into the halo centres. 

Friends-of-Friends (\textsc{fof}) groups are identified in the simulation using a
linking length of 0.2 times the mean inter particle separation \citep{Davis1985}. Sub-haloes are 
found starting from \textsc{fof} groups using {\sc subfind} \citep{Springel2001}. 
To identify the halo centre, we first find the most massive sub-halo within each 
\textsc{fof} halo, then sort all particles within the sub-halo according to their total energy. 
The most bound particle, i.e. the one with the lowest total energy is treated as the halo 
centre. Note that this centre is usually different from the minimum of the potential of the group. 
Halo masses are defined as the mass around the halo centre and within the radius $r_{200}$, where $r_{200}$ is the radius within which the mean density is 200 times of the 
critical density of the Universe.

We use the {\sc Gadget-3} \citep{Springel2008} code to evaluate the peculiar potentials for all particles. 
 In each cluster we define the potential of the cluster centre
to be the mean of that of particles within a core radius of 3 kpc/$h$ of the cluster centre.
Changing the size of the core radius simply shifts the profiles of
$\Phi_{\rm c}-\Phi(r)$
up and down, but does not alter their shapes.

To estimate the stacked potential profiles for the case of isotropic averaging, we insert massless test particles uniformly 
distributed on spherical shells around each halo centre.
We use {\sc Healpix} \citep{Gorski2005} to generate 3072 pixel coordinates over 4$\pi$ stradians. The
mean spacing of pixels is about 3.6 degrees. Along the radial direction,
20 spherical shells are distributed logarithmically per decade, starting at 0.01
Mpc/$h$ from the halo centres out to 30 Mpc/$h$. Convergence of the results has been tested in the radial and tangential directions by doubling the sampling rate along both directions 
respectively, and we have found no noticeable change in our results. 
The $z=0$ snapshot is used for our study.

\section{Results from 3D profiles}
Fig.~\ref{Fig:Individual} gives two examples of clusters from the Millennium simulation.
The matter distribution of these clusters is far from spherical, especially at large radii 
from the cluster centres. There are filamentary structures and in some cases, massive neighbouring
clusters within a 10~Mpc/$h$ radius. The potentials projected along one
major axis of the simulation box clearly illustrate the complex morphology of the
potential within the 10 Mpc/$h$ radius regions.
The main central haloes are associated with the deep potential wells. The bottom panels show the corresponding mass-weighted potential profiles $\Phi^{\rm obs}$  of Eq.~(\ref{Eq:obs})
which are similar to what will be observed (except that this is in real space), and the spherically averaged profiles, $\Phi^{\rm iso}$,
evaluated using the isotropic averaging of Eq.~(\ref{Eq:iso}). 

The main halo shown on the left is relatively massive, $10^{15}{\rm M_{\odot}}/h$. 
The shape of the potential well may seem symmetric close to the bottom of the minimum (middle-left). However it is not, as it can be
seen that the colour distribution is clearly not symmetric. 
This indicates variation in the projected mass density. 
It is consistent with the picture in the top-left panel, where a bar-shaped core is visible close to the centre. 
Along the direction of elongation near the core of the cluster, 
the potential values are slightly deeper than those along the perpendicular short axis. 
Seen in 3D, the potential well looks like a valley floor, where more mass is concentrated along the valley. At small radii, 
this causes the mass-weighted potential profile $\Phi^{\rm obs}$ to be shallower than the spherically averaged profile, $\Phi^{\rm iso}$. 
One can see the corresponding kink 
within 1~Mpc/$h$ in the lowermost left panel.

The neighbouring structures also
induce local potential minima. For relatively small neighbours, the
neighbouring potential minima are shallower than the central one. 
This is the case for the left hand figure. 
The non-spherical nature of the main halo and the presence of neighbours make the mass weighted
potential profiles $\Phi^{\rm obs}$ shallower than
the case of spherical averaging $\Phi^{\rm iso}$, as seen in the bottom panels. 

As the mass of the main halo gets smaller, the chance of having a comparably massive
neighbouring structure increases. In some cases a neighbouring
cluster can be even more massive than the main central halo,
as shown by the example in the right-hand panels of
Fig.~\ref{Fig:Individual}. Here the mass of the main halo is $10^{14}{\rm M_{\odot}}/h$. 
At about 5 Mpc/$h$ from the chosen main halo centre, one of neighbouring haloes has a deeper potential minimum 
than the main halo. The very massive neighbours cause strong biases of the potential profile (bottom-right panel). 
Note that the neighbouring systems shown in Fig.~\ref{Fig:Individual} are not sub-structures of the main halo.
They are essentially correlated large-scale structures outside the virial radius of the main halo.

\subsection{The stacked potential profiles}

With the individual observed potential profiles $\Phi^{\rm obs}$ always being biased low compared to the 
spherically averaged $\Phi^{\rm iso}$, it is clear that the stacked profiles can not be given by averaging the $\Phi^{\rm iso}$ profiles, even if the stacked cluster system is perfectly spherically symmetric. 
Results for two different halo mass ranges are shown in Fig.~\ref{Fig:Stack}. As expected, the stacked profiles of
$\bar \Phi^{\rm obs}$ are systematically lower, resulting in less negative values of $\Phi_{\rm c}-\bar \Phi^{\rm obs}$ in Fig.~\ref{Fig:Stack} (solid lines) than the corresponding 
spherically averaged profiles, $\Phi_{\rm c}-\bar \Phi^{\rm iso}$ (dotted lines). This indicates that the blueshifts of the surrounding galaxies relative to the central BCGs will be smaller than predicted
by the assumption of spherical symmetry.
We find the absolute difference between $\Phi^{\rm obs}$ and $\Phi^{\rm iso}$ for $M>10^{14}$~M$_{\odot}/h$ can be well approximated 
by a linear function when the radii are rescaled by $r_{200}$, i.e. 
$\Delta \Phi/c\approx -0.25r/r_{200}$~km/s. This approximation is shown by the solid lines in the lower panels of Fig.~\ref{Fig:Stack}.
This approximated relation also holds for the case of $M>2\times 10^{13}$~M$_{\odot}/h$ at $r<5$~Mpc/$h$.
We find that this approximation holds for a wide range of minimum halo masses 
${\rm M_{min}}$, from $10^{13}$ to $10^{15}{\rm M_{\odot}}/h$. 
In terms of fractional differences, $(\bar \Phi^{\rm obs}-\bar \Phi^{\rm iso})/\bar \Phi^{\rm obs}$, these are stronger when ${\rm M_{min}}$ is small. 
For ${\rm M_{\rm min}}=2\times 10^{13}{\rm M_{\odot}}/h$ shown in the left panel of Fig.~\ref{Fig:Stack}, 
the bias is approximately 60\% at $r>5$ Mpc/$h$. 
For ${\rm M_{\rm min}}=10^{14}~{\rm M_{\odot}}/h$ (right), the bias varies from a few percent to more than 20\%.
We also find that in an extreme case when clusters with relatively low halo mass 
($1\times 10^{13}<M<2\times 10^{13}{\rm M_{\odot}}/h$) are used, the mass weighted potential 
profiles are very close to zero at most scales due to the presence of neighbouring structures.

It is noticeable that both $\Phi_{\rm c} -\bar \Phi^{\rm obs}$ and $\Phi_{\rm c} -\bar \Phi^{\rm iso}$ have 
troughs at approximately 2~Mpc/$h$ caused by the presence of the secondary potential wells at radii greater than 2~Mpc/$h$, which cause $\Phi_{\rm c} -\bar \Phi^{\rm obs}$ to become less negative at $r>$2~Mpc/$h$. The rises on the right hand side of the 
troughs seen in Fig.~\ref{Fig:Stack} are more pronounced
for $\Phi_{\rm c} -\bar \Phi^{\rm obs}$ than for $\Phi_{\rm c} -\bar \Phi^{\rm iso}$ as mass weighting gives more
weight to the secondary potential wells.  
The troughs are also stronger for lower values of ${\rm M_{\rm min}}$ as the chance of having more massive neighbours is greater. 
%It might seem puzzling that there should be a trough even for the spherical averaged case. 
%This is because in this case, the $\bar \Phi^{\rm iso}$ value at each $r$ is contributed 
%mostly by the more numerous low-mass haloes. The estimate of $\bar \Phi^{\rm iso}$ is analogous to 
%averaging the two dashed curves in the bottom panels of Fig.~\ref{Fig:Individual}, but with more weight being given to the curve on the right. A trough in the averaged curve is naturally expected.
Note that the troughs are not seen in previous models in the literature, e.g. \citep{Wojtak2011}. 
One may suspect that this might be due to the fact that the profiles we show here are from 3D averaging.
We will show in the next section that even when quantified by projected distances, the troughs in the profiles remain. So projection effects are not the explanation for the absence of the troughs.

Note that the biases of the dotted lines with respect to the solid lines  
are purely due to the assumption of spherical symmetry. For completeness, 
we also compare them with the most simplistic case where the weighted contributions 
from each of the individual haloes at each $r$ are assumed to be equal. i.e.
$\Phi^{\rm idl}$ of Eq.~(\ref{Eq:idl}). These results are shown by the dashed lines in Fig.~\ref{Fig:Stack}. 
No trough is seen and the profiles are smooth and monotonic. 
The biases for this case of $\bar \Phi^{\rm idl}$ versus $\bar \Phi^{\rm obs}$, defined as ($\bar \Phi^{\rm obs}-\bar \Phi^{\rm idl})/\bar \Phi^{\rm obs}$, are 40\% (left) and 20\% (right) at their maxima, as shown by the dashed curves in the bottom panels of Fig.~\ref{Fig:Stack}.

In summary, haloes are in general ellipsoidal rather than spherical. Within the virial radius of a halo, there is more mass concentrated along the long axis of the halo. The higher mass concentration generates deeper potential valleys along the major axis. The mass weighted potentials are therefore higher than the case of spherical averaging. Outside the virial radius of the halo, the matter distribution is even further from being spherical distributed. The filamentary structures and neighbouring haloes embedded in them create deep secondary potential minima. These tend to decrease the potential difference with respect to the cluster centre. The difference for the potential profiles between mass weighting versus spherical averaging is comparable to the model differences between the predicted gravitational
redshifts for some modified gravity theories and General Relativity \citep{Wojtak2011}.
The biases, if not accounted for, may confuse the interpretation of the observed signal. However, we will show in the next section that the picture we have presented so far will change significantly when observing particles/galaxies in velocity space. Also, the predicted signal will 
be altered by the other terms arising from the treatment of the past light cone. 

%$\bullet$ Discuss the size of the core matters for the amplitudes of the
%potentials and forces.

\begin{figure*}
\begin{center}
\scalebox{0.39}{
\includegraphics[angle=0]{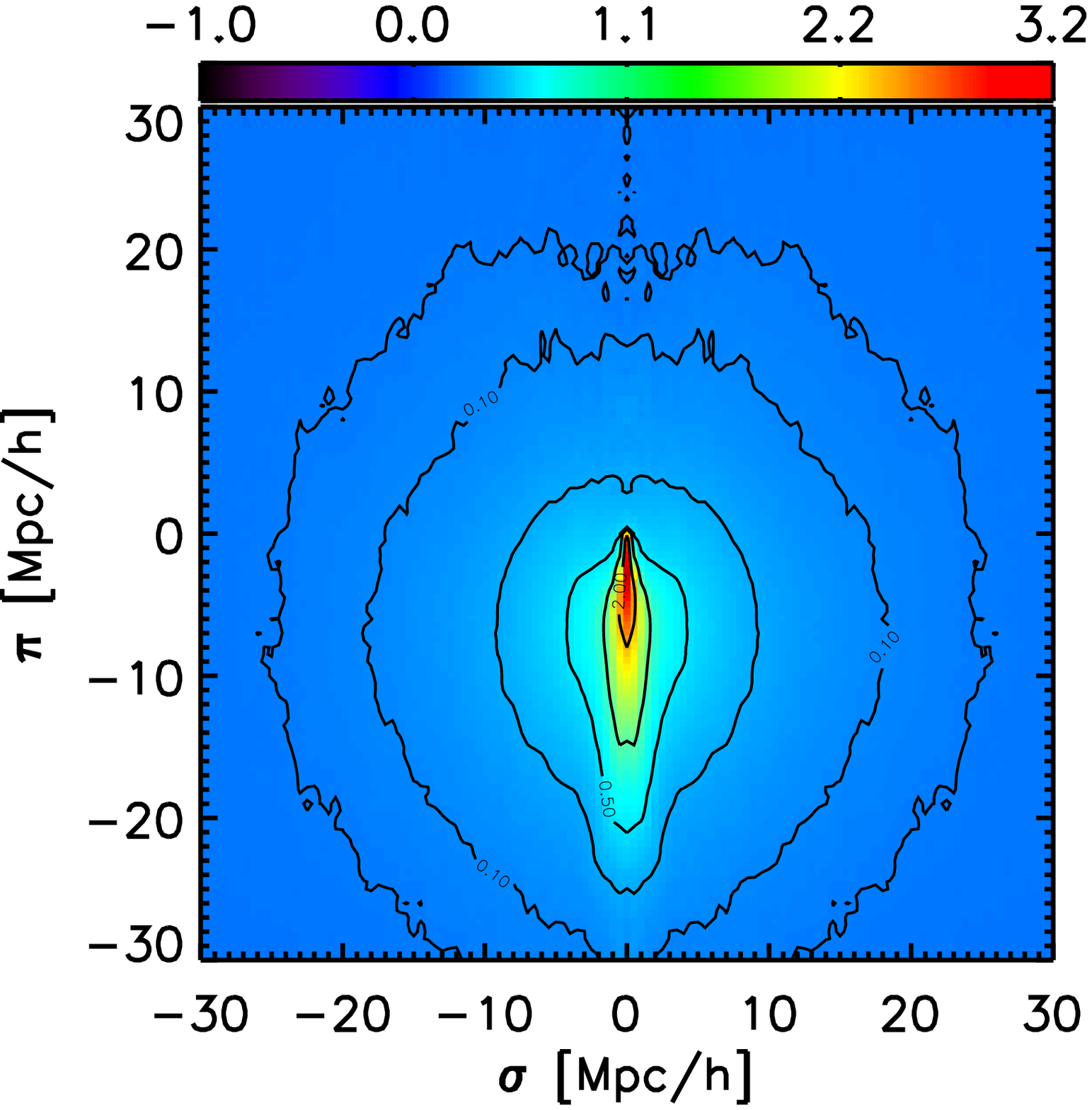}}
\hspace{-2.5cm}
\vspace{-0.3cm}
\scalebox{0.39}{
\includegraphics[angle=0]{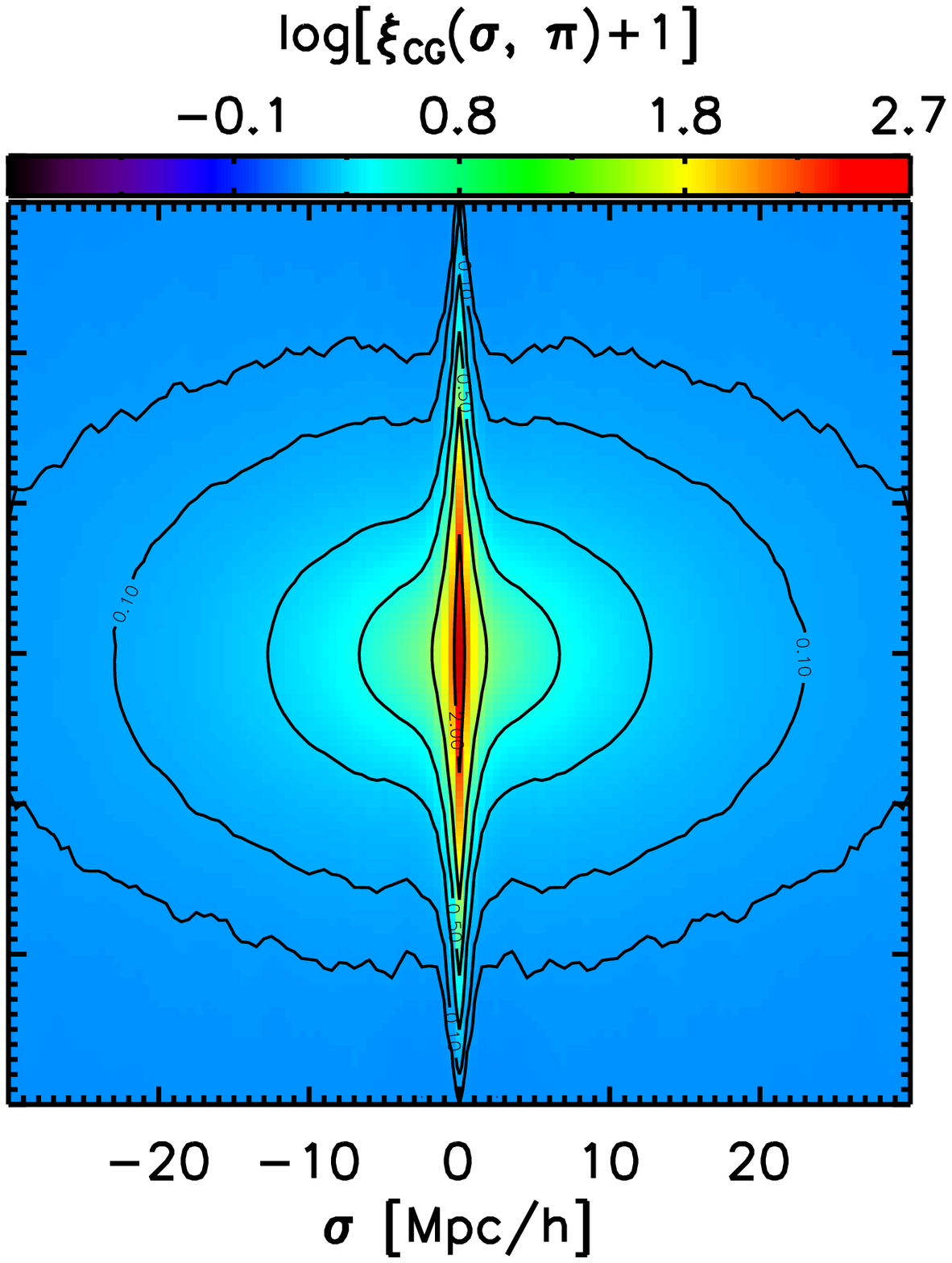}}
\hspace{-2.5cm}
\vspace{-0.3cm}
\scalebox{0.39}{
\includegraphics[angle=0]{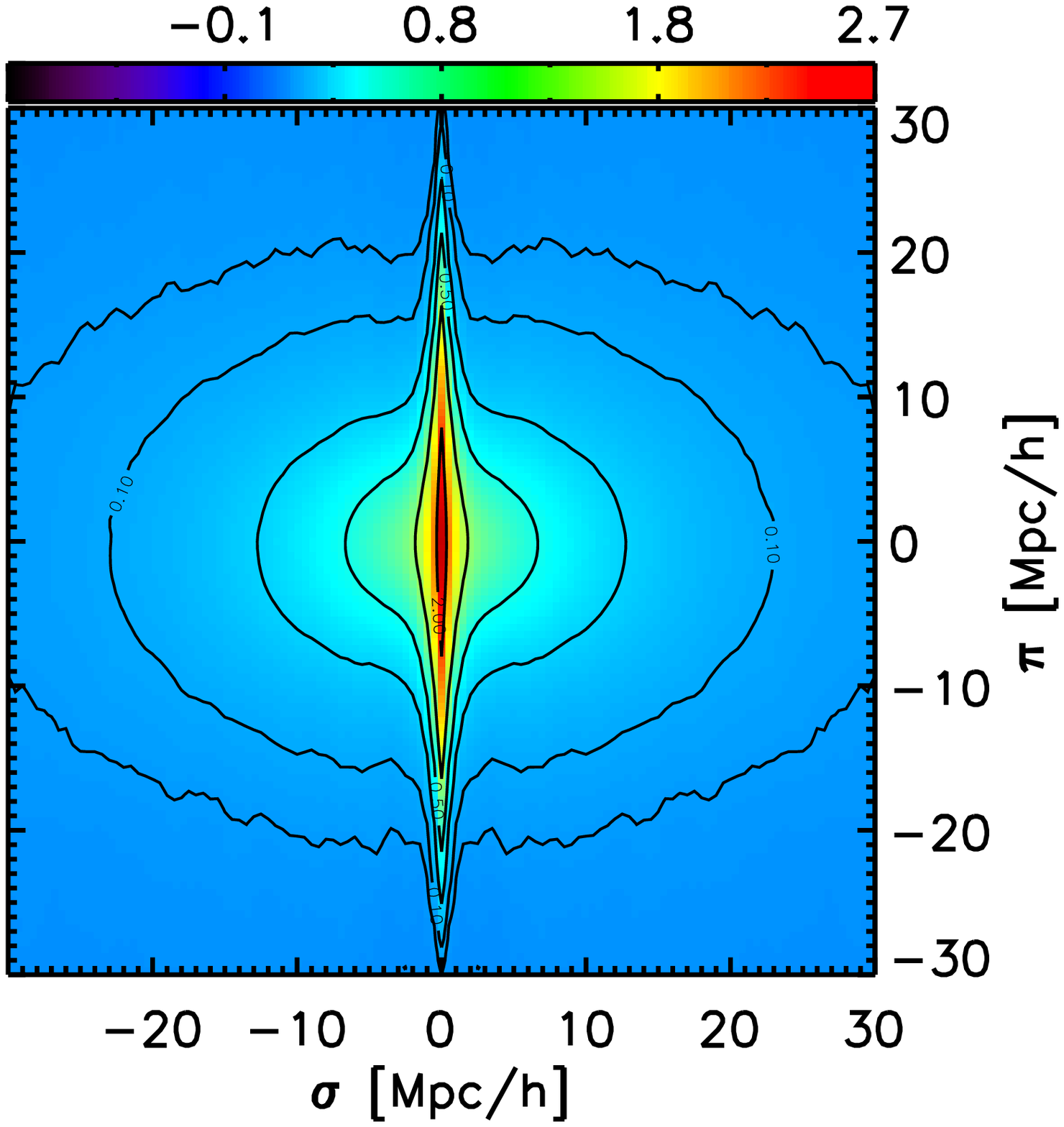}
}
%%%%row 2
\scalebox{0.38}{
\hspace{-3.12 cm}
\includegraphics[angle=0]{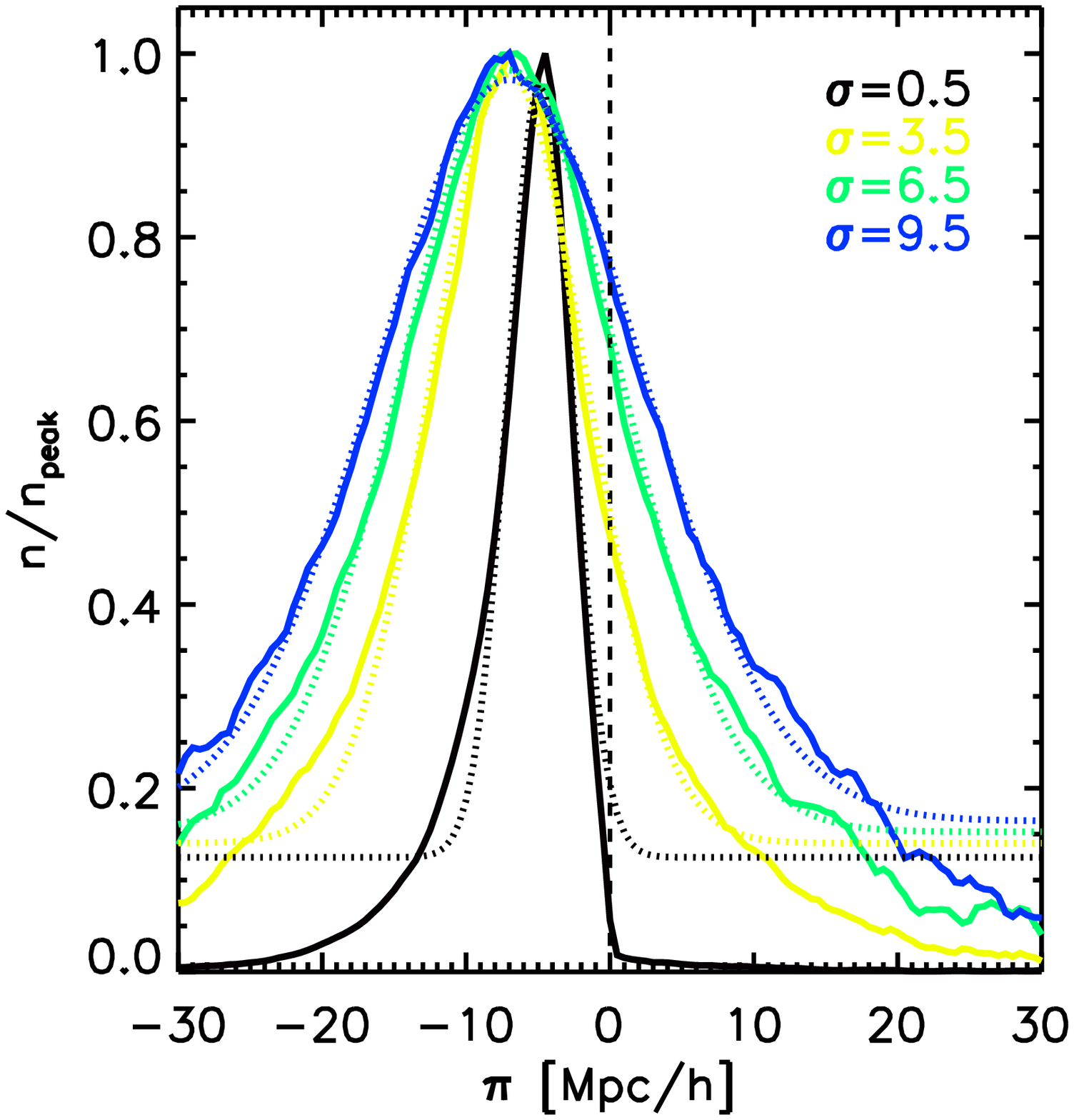}}
\hspace{-2.3cm}
\vspace{-0.3cm}
\scalebox{0.38}{
\includegraphics[angle=0]{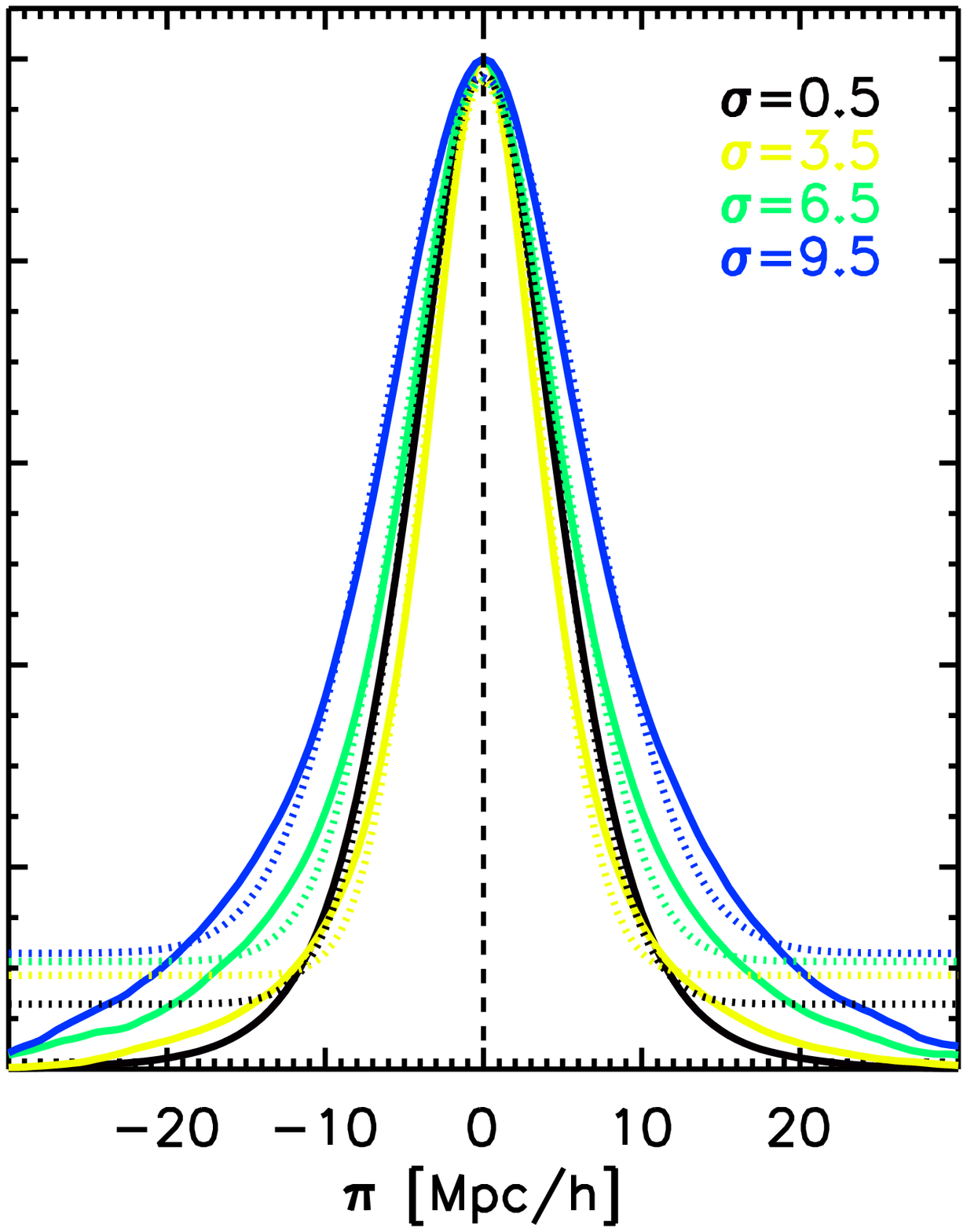}}
\hspace{-2.3cm}
\vspace{-0.3cm}
\scalebox{0.38}{
\includegraphics[angle=0]{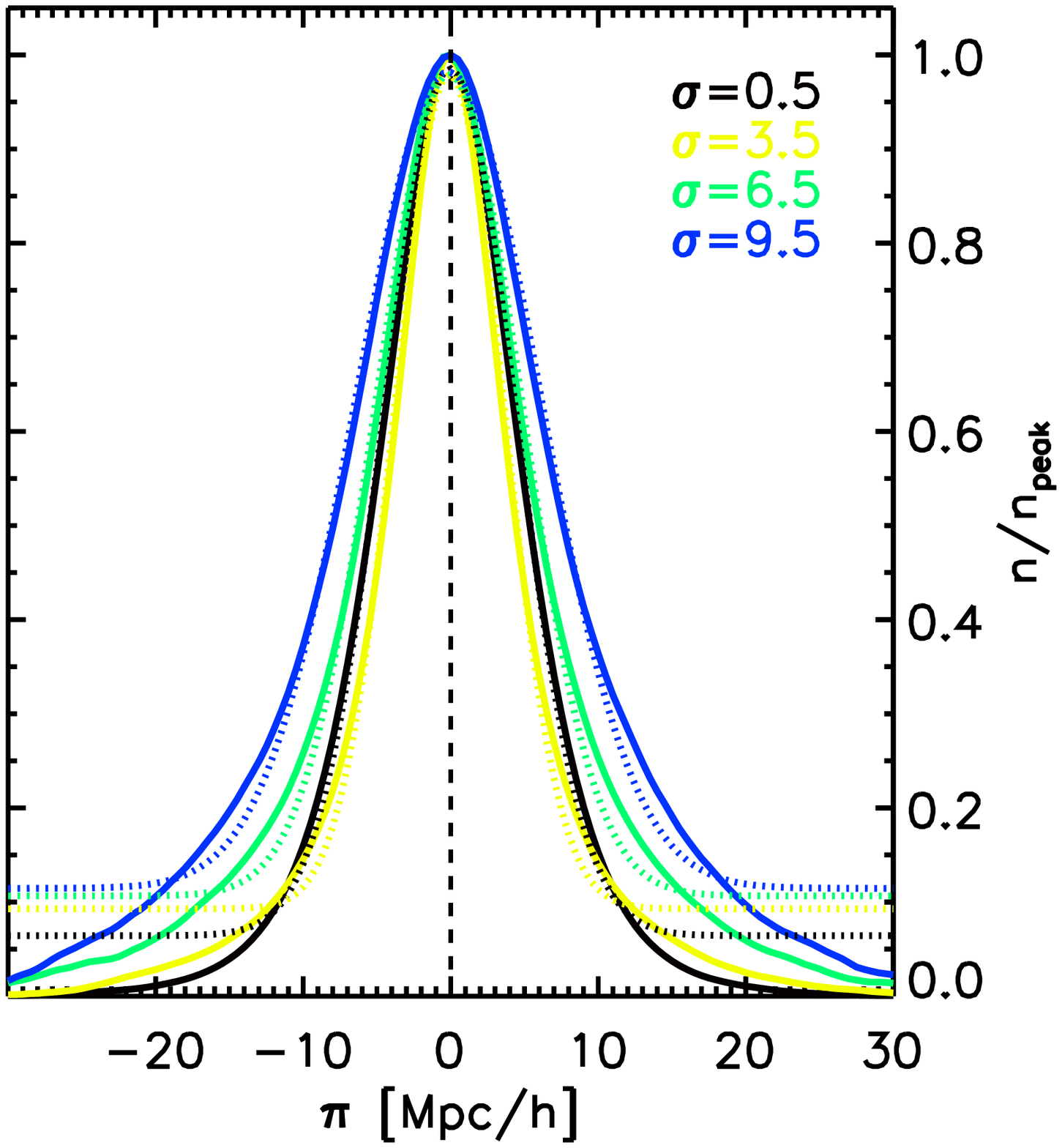}}
%%%%row 3
\scalebox{0.38}{
\hspace{-3.12 cm}
\includegraphics[angle=0]{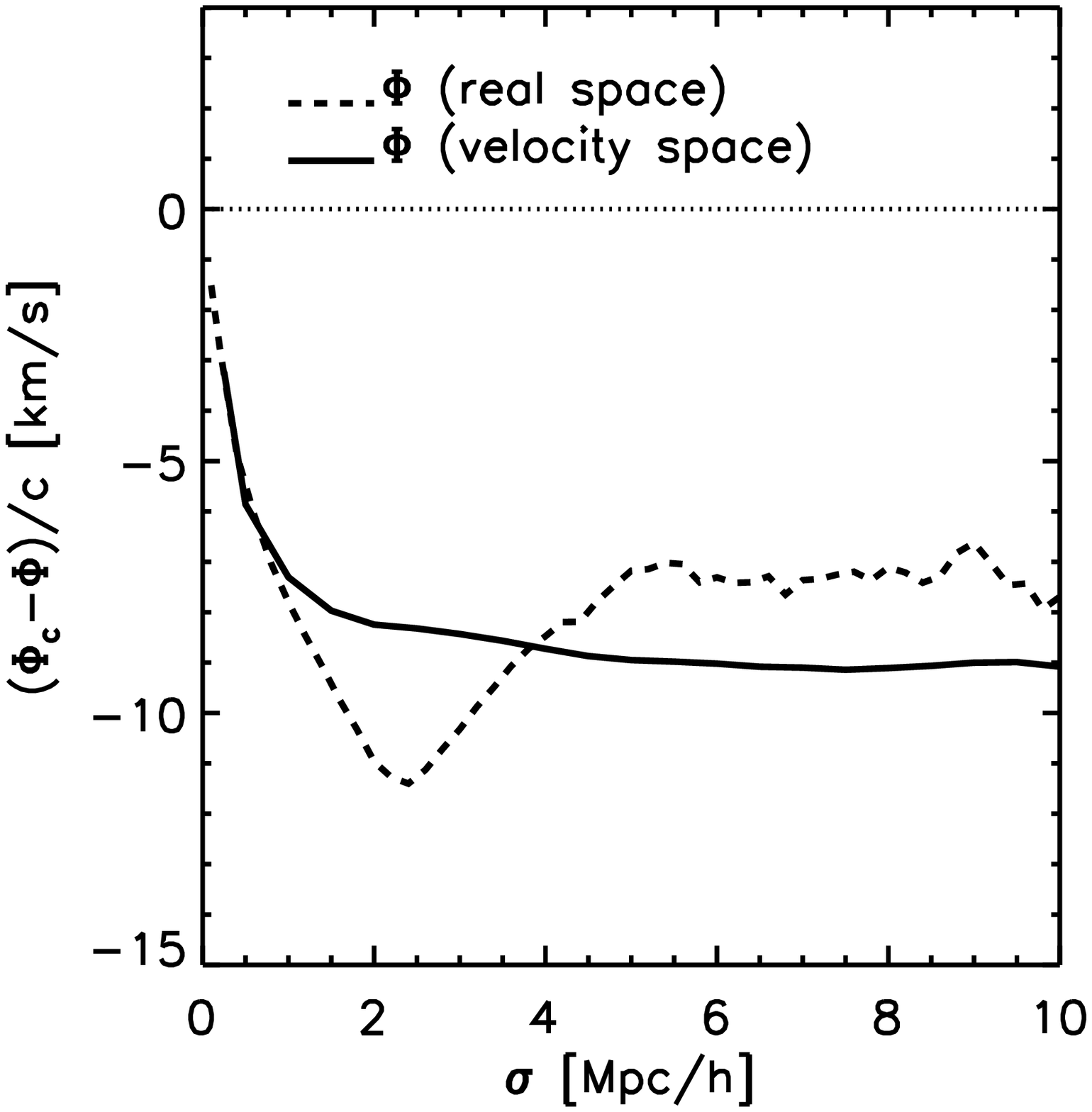}}
\hspace{-2.3cm}
\scalebox{0.38}{
\includegraphics[angle=0]{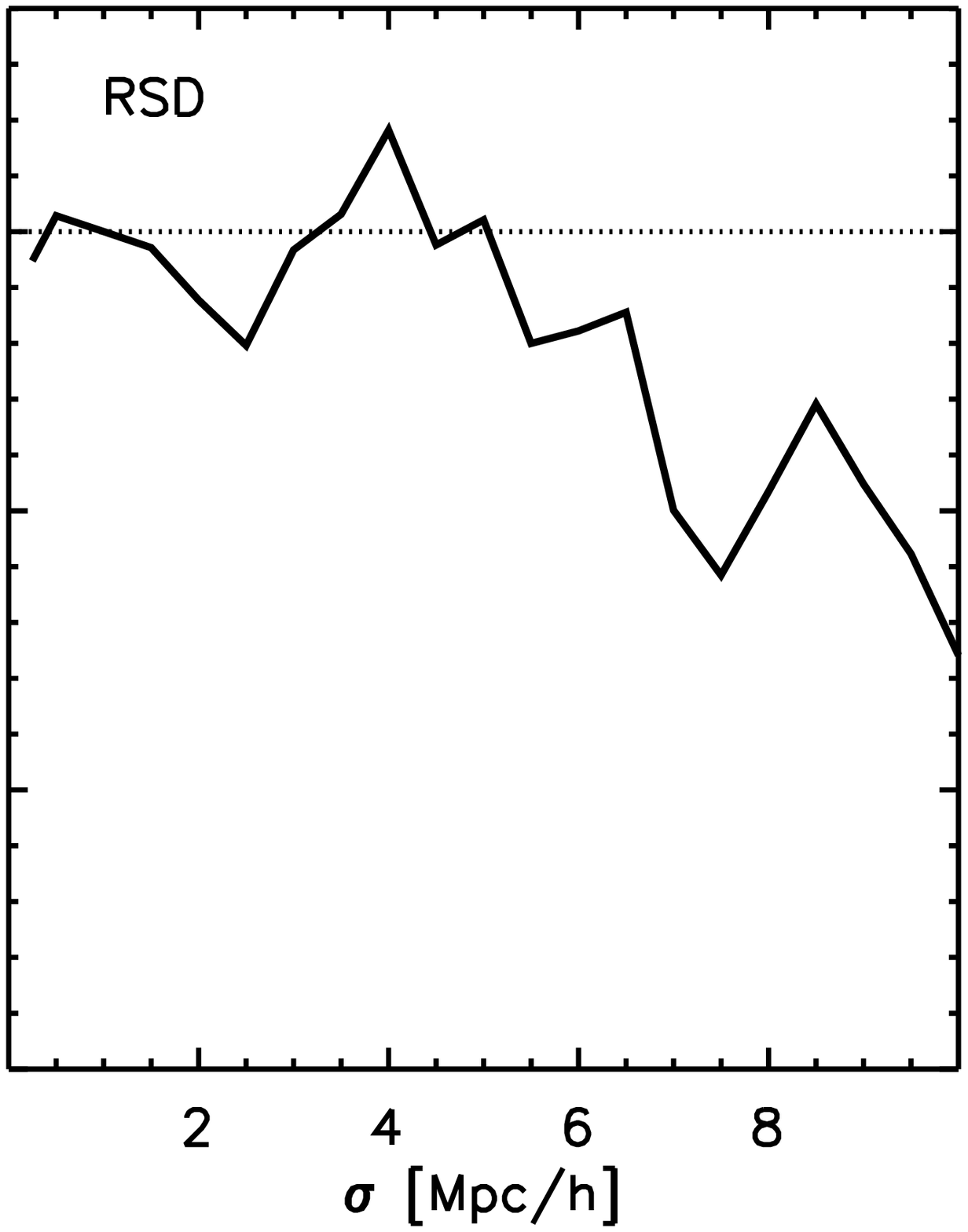}}
\hspace{-2.3cm}
\scalebox{0.38}{
\includegraphics[angle=0]{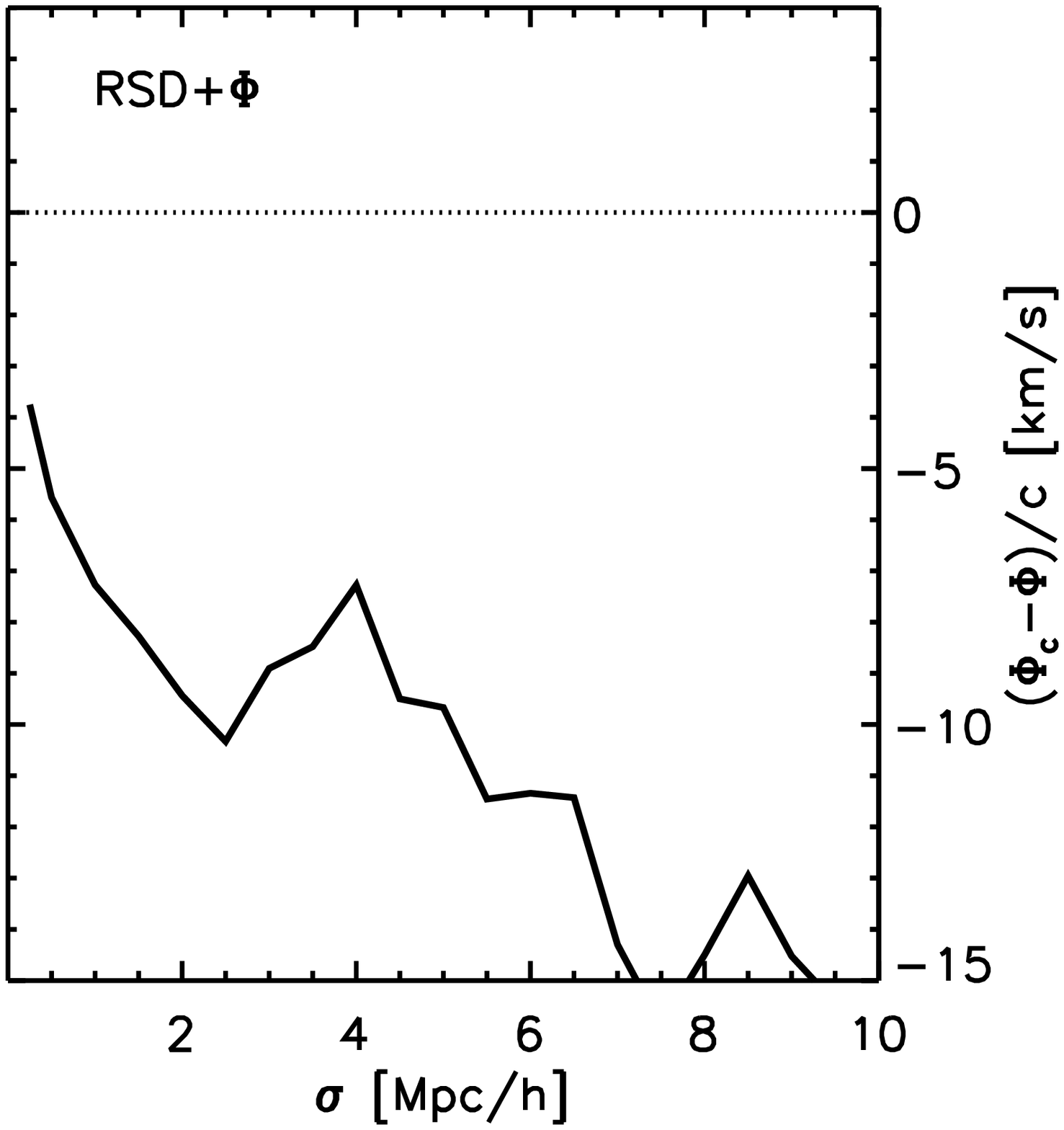}
}
\caption{{\it Top row}: the cluster-mass correlation functions in gravitational redshift space (left), in peculiar velocity space (RSD) (middle) 
and the sum of the two (right) for haloes with the mass $M>5\times 10^{13} {\rm M_{\odot}}/h$. In the left-hand panel, no peculiar velocities are added and the gravitational redshift signal has 
been amplified by 100 times for better visualisation. In the right-hand panel, the gravitational redshift distortion is much smaller than that from peculiar velocities and it is difficult to see the difference it produces relative to the middle-panel. {\it Middle row}: Examples of histograms of the particle distributions (from the top panels) along the line-of-sight direction, $\pi$, at different projected 
distances, $\sigma$, from the cluster centres. Dotted lines are the best fit models [Eq.~(\ref{EqGaussian})] to the solid lines. The offsets of the fitted peaks from the 
centre are interpreted as the gravitational redshift signal. {\it Bottom row}: The best-fit values for the offsets from the centre shown in the middle row. Subtracting the curve on the right from the one in the middle yields the solid curve on the left, which is the velocity space version of the GRedshift 
signal. The dashed curve on the left is the real space version. The non-zero values in the middle panel are due to sample variance.} 
\label{FigRSD}
\end{center}
\end{figure*}

\section{the ultimate redshift space distortion}

Results from the previous sections use 3D spatial information for clusters in real space, which is useful for understanding the physics. In this section, we take one step closer to the observations by placing the simulated clusters in velocity space and including the past light cone effects. We will quantify each term on the RHS of Eq.~(\ref{Eq:full2}) and compare them to the GRedshift signal. This can be achieved by computing the cluster-galaxy correlation function (CGCF) and identifying asymmetries in the correlation function along the line of sight. In the following, starting with the example of gravitational redshift, we will demonstrate how the asymmetric features associated with the final three terms of Eq.~(\ref{Eq:full2}) are recovered and how they differ between real space and velocity space.

\subsection{Gravitational redshift in real space}
As an intermediate step and for the purpose of comparison, we first measure the impact of GRedshift in the CGCF in real space, i.e. without the
perturbing effect of peculiar velocities.
In the simulations, we turn off the velocity term in Eq.~(\ref{Eq:full2}) so that the particles are displaced only by gravitational redshifts. The 
cluster-galaxy (or halo centre-particle) cross-correlation function (CGCF), $\xi(\sigma, \pi)$ shown in the top-left panel 
of Fig.~\ref{FigRSD}, is computed using all particles. 
For this figure, the amplitudes of the $\Phi$'s have been artificially boosted by a factor of $100$ to aid visualisation. 
The cluster centres are 
redshifted most as they are at the bottom of the potential wells. The amount of gravitational redshift decreases as the distance relative to the cluster centre increases. 
Particles are therefore preferentially shifted towards the observer
at large separations both  in front of and behind the cluster centre. This effect decreases with  projected distance, causing the inverted-candle-flame shape for the correlation function. 
Relative to the cluster centre, the rest of the particles are blueshifted with an amplitude that generally increases with
projected distance, but with the exception of when the impact of neighbouring groups or clusters is significant.

The amplitudes of the GRedshifts are measured by locating the peak of the particle distribution function (PDF) at 
a given projected distance 
$\sigma$ from the stacked cluster centre. The peaks are located by fitting the PDFs with a Gaussian plus a constant: 
\begin{equation}
\label{EqGaussian}
f(y)=A+B\exp{[-(y+\hat\Phi^{\rm obs})^2/C]},
\end{equation}
where A, B, C and $\hat{\Phi}^{\rm obs}$ are free parameters, and $\hat{\Phi}^{\rm obs}$ is the parameter of interest. 
The middle-left panel shows examples of the measured PDFs (solid) and the best fit results (dotted). The poor agreement between the solid and the dotted lines away from the peaks does not affect our results, as we are only interested in the locations of peaks.

The dashed curve in the bottom-left panel is the recovered amplitude of $\Phi^{\rm obs}$. A bump at $\sim$2~Mpc/$h$ caused by neighbouring clusters is clearly seen. This is consistent with what we found when measuring the potential profiles directly (in real space) as shown in Fig.~\ref{Fig:Stack}.

Note that in practice, with realistic amplitudes of the $\Phi$'s, it is nontrivial to recover the asymmetry of the correlation function purely due to gravitational redshift. Sample variance will strongly affect the measurement. Although in principle, it can be beaten down by using very large samples, we are limited by the size of the simulation. It is therefore necessary to employ another technique to suppress the variance.
Boosting $\Phi$ helps to illustrate the effect,
 but we think that this method is not ideal as sample variance is not completely eliminated. Instead, in the next subsection
we adopt another method to eliminate sample variance. This
new method is found to be robust regardless of the amplitude of the signal. 

\subsection{Gravitational redshift in velocity space}
\subsubsection{signal-to-noise}

In reality measurements are made in redshift space including $\Phi$ etc, but from Eq.~(\ref{Eq:full2}), it is obvious that the peculiar velocity term is the most dominant
and so to a good approximation
all other quantities are measured in velocity space. The gravitational redshift signal is of the order of 10~km/s for haloes with masses above $5\times 10^{13}{\rm M_{\odot}}/h$. This is at least one order of magnitude smaller than the peculiar velocity dispersion of clusters $\sigma_v$. Suppose we have $\sigma_v=400$~km/s, and we want to achieve a $3\sigma$ detection of the gravitational redshift. We will need to stack of order 14,000 clusters. We list the estimated signal-to-noise of the GRedshift signal for a few cluster samples with different velocity dispersions in Table~1. 
 The signal-to-noise increases with decreasing $\sigma_v$ or the minimum halo mass ${\rm M_{\rm min}}$.
A 5$\sigma$ detection can be achieved with a group sample of $M>5\times 10^{13}\,{\rm M_{\odot}}/h$ or $\sigma_v>400$~km/s in a volume of 1~(Gpc/$h$)$^3$.
\begin{table}
\centering
\caption{Estimated signal-to-noise for the gravitational redshift signal from stacked clusters 
from the Millennium simulation. The $\Phi$ values are taken from the minima of the stacked profiles.} 
\bigskip
\begin{tabular}{lccccccc}
\hline
${\rm M_{\rm min}}$ & $\sigma_{v}$ & $N_{\rm halo}$ & $\Phi$ & S/N for  & S/N for \\ 
$[{\rm M_{\odot}}/h]$  &  [km/s] &  & [km/s] &  1/8 (Gpc$/h)^3$ & 1 (Gpc$/h)^3$ \\
       \hline 
$1\times 10^{13}$ &  240 &  35300 & 3 & 2.3 & 6.6\\
$5\times 10^{13}$ & 400  &  5283 & 10  & 1.8  & 5.1 \\
$1.6 \times 10^{14}$  & 600  &  1000 & 15   & 0.7  & 2.1\\
$3.5 \times 10^{14}$  & 800 & 180 & 22  &  0.3  & 1.0 \\
\hline
\end{tabular}
\label{tab2}
\end{table}
Within the dynamical range of haloes from Table 1, $\Phi$ versus $\sigma_{v}$ can be fitted by a linear function $\Phi/c=(\sigma_v/30-5)$~km/s. Or $\Phi/c=[18{\rm (M_{\rm min}}/10^{13})^{1/3}-5]$~km/s.

%{\bf (For the first case, you have 5283 clusters and when you do the 3-axis stacking, you've got 15000 clusters, which is about OK. but for the orange colour case, it is not possible from your sample, this is perhaps why  in Fig. 14. The black curves are much well behaved on the left than on the right.)} 
In the Millennium simulation, we have 5283 clusters with 
$M>5\times 10^{13}\,{\rm M_{\odot}}/h$. This is only 1/3 of the halo number needed achieve a $3\sigma$ detection. To increase the sample without using a larger box-size simulation, we view the simulation along its three principal axes. This effectively increase the number of haloes by a factor of 3. 
\subsubsection{redshift-space distortions only}
To highlight the impact of sample variance, we show in the middle column of Fig.~\ref{FigRSD} results without gravitational redshift, and with peculiar velocity distortions only. This is the case of conventional RSD. In principle, without sample variance, no asymmetry along the line of sight is expected. When we follow the same procedure by using Eq.~(\ref{EqGaussian}) to fit for $\Phi^{\rm obs}$, it should be zero. However, from the bottom-middle panel, we see the measured $\Phi^{\rm obs}$ fluctuates around zero at the level of a few km/s. The expected dispersion of 
$\Phi^{\rm obs}$ estimated from the velocity dispersion and size of this sample is approximately 3~km/s. The amplitude of fluctuations are consistent with sample variance.
%Also, it is important to note that in real observations, the measurement is effectively conducted in velocity space because the effect of peculiar velocity distortions is dominant. This will cause subtile differences for the recovered GRedshift signal compared to that in real space. 
\subsubsection{Gravitational redshift + RSD}

\begin{figure*}
\begin{center}
\advance\leftskip -1.0cm
\scalebox{0.5}{
\includegraphics[angle=0]{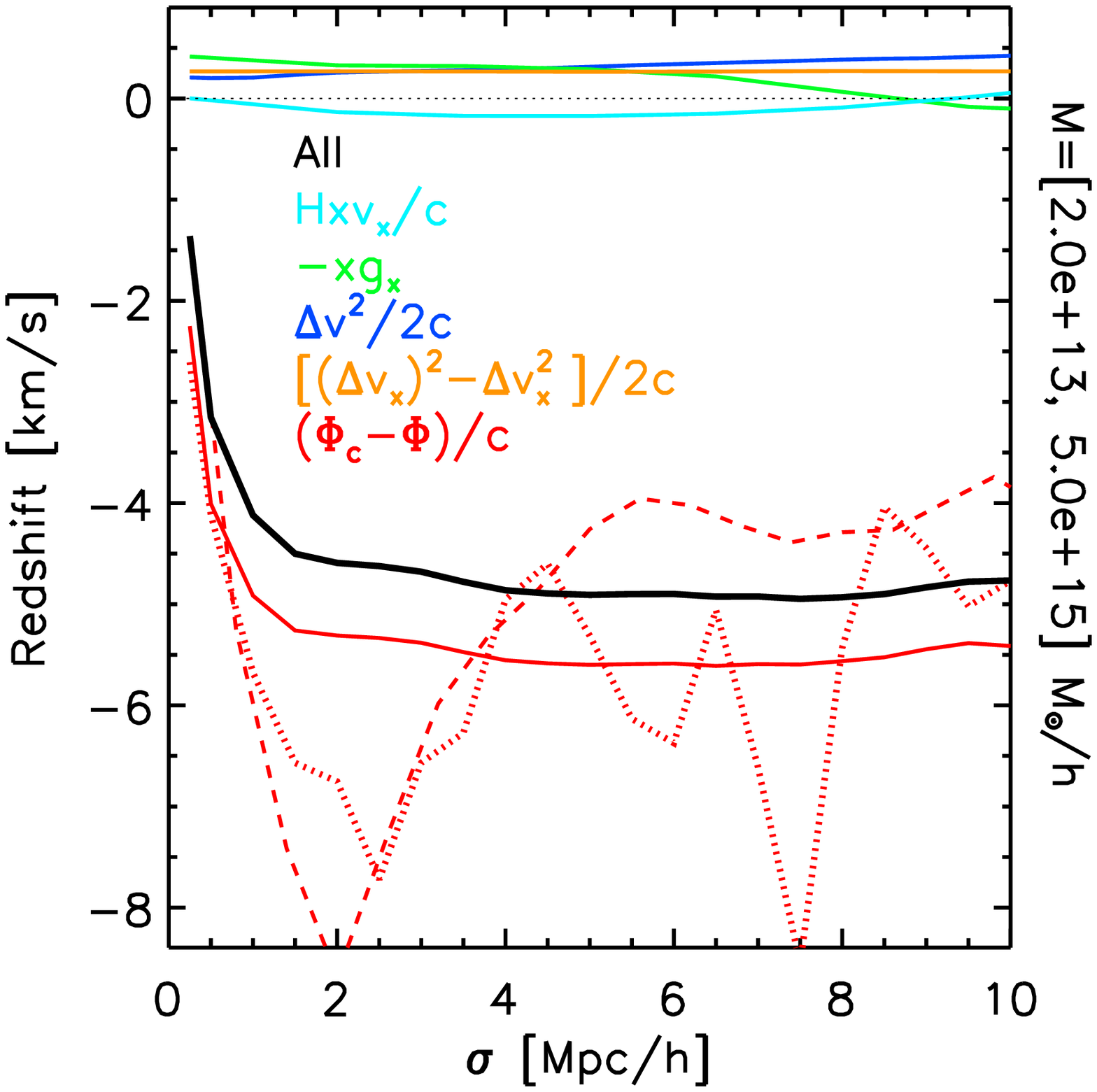}
\includegraphics[angle=0]{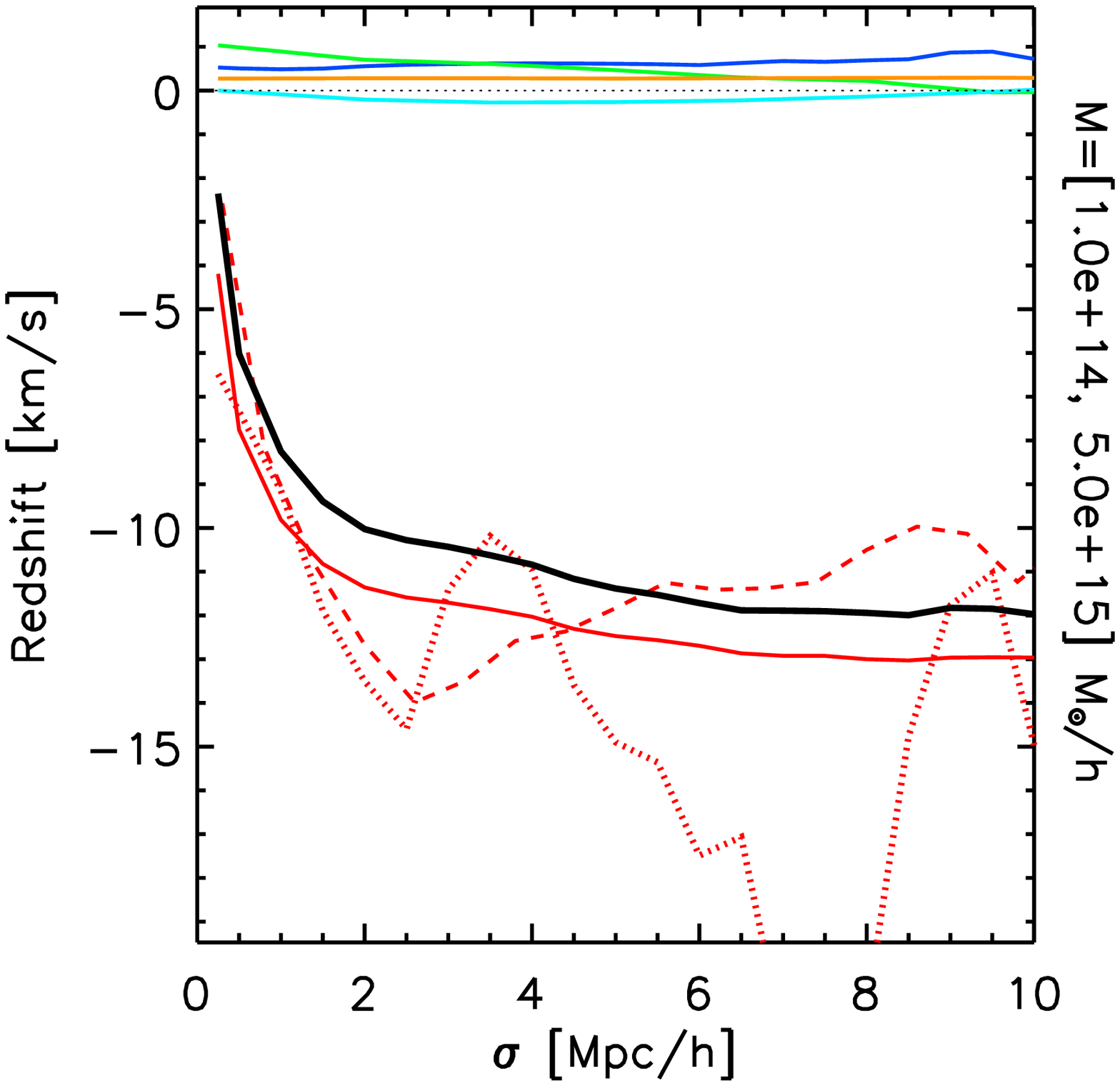}}
\caption{The dependence of the redshift offset of the peak of
cluster-mass cross-correlation function on transverse separation, $\sigma$. The two panels are for different halo mass ranges
as labelled on their right hand axes.
$(\Phi_{\rm c}-\Phi)/c$ (red solid) is the idealised gravitational redshift signal {\it in velocity space}. The red dashed curves are their {\it real space} versions. They are comparable to the solid curves in Fig.\ref{Fig:Stack}. 
Blue and orange represents quantities related to the the special relativistic effect. Green and cyan curves represents the other two terms arising from the effects of the past light cone. The black curves are the sum of all the terms. The actual measurements from simulations including sample variance are shown by the red-dotted curves. }
\label{Fig:AllTerms}
\end{center}
\end{figure*}

\begin{figure*}
\begin{center}
\advance\leftskip -1.0cm
\scalebox{0.5}{
\includegraphics[angle=0]{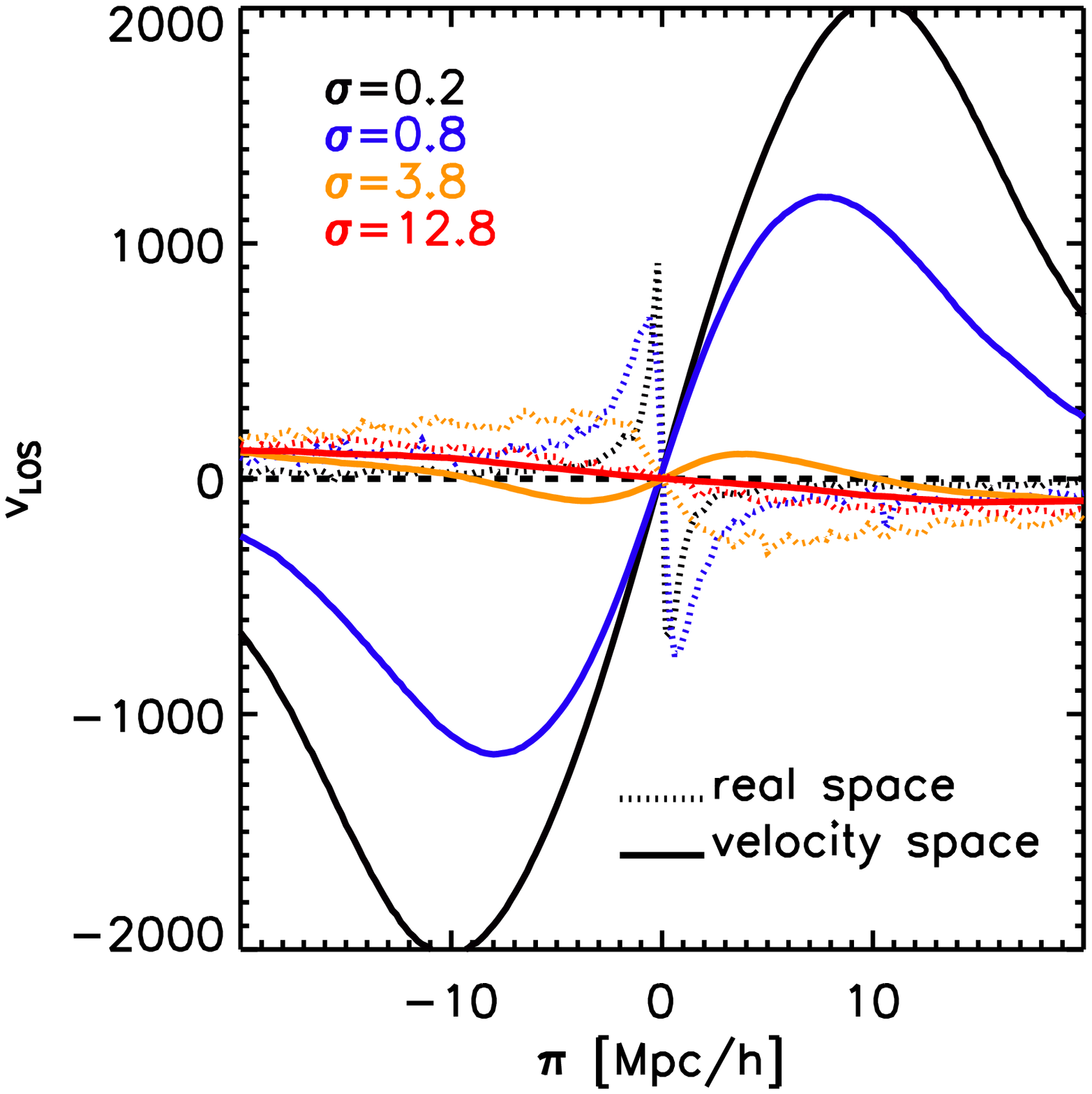}
\includegraphics[angle=0]{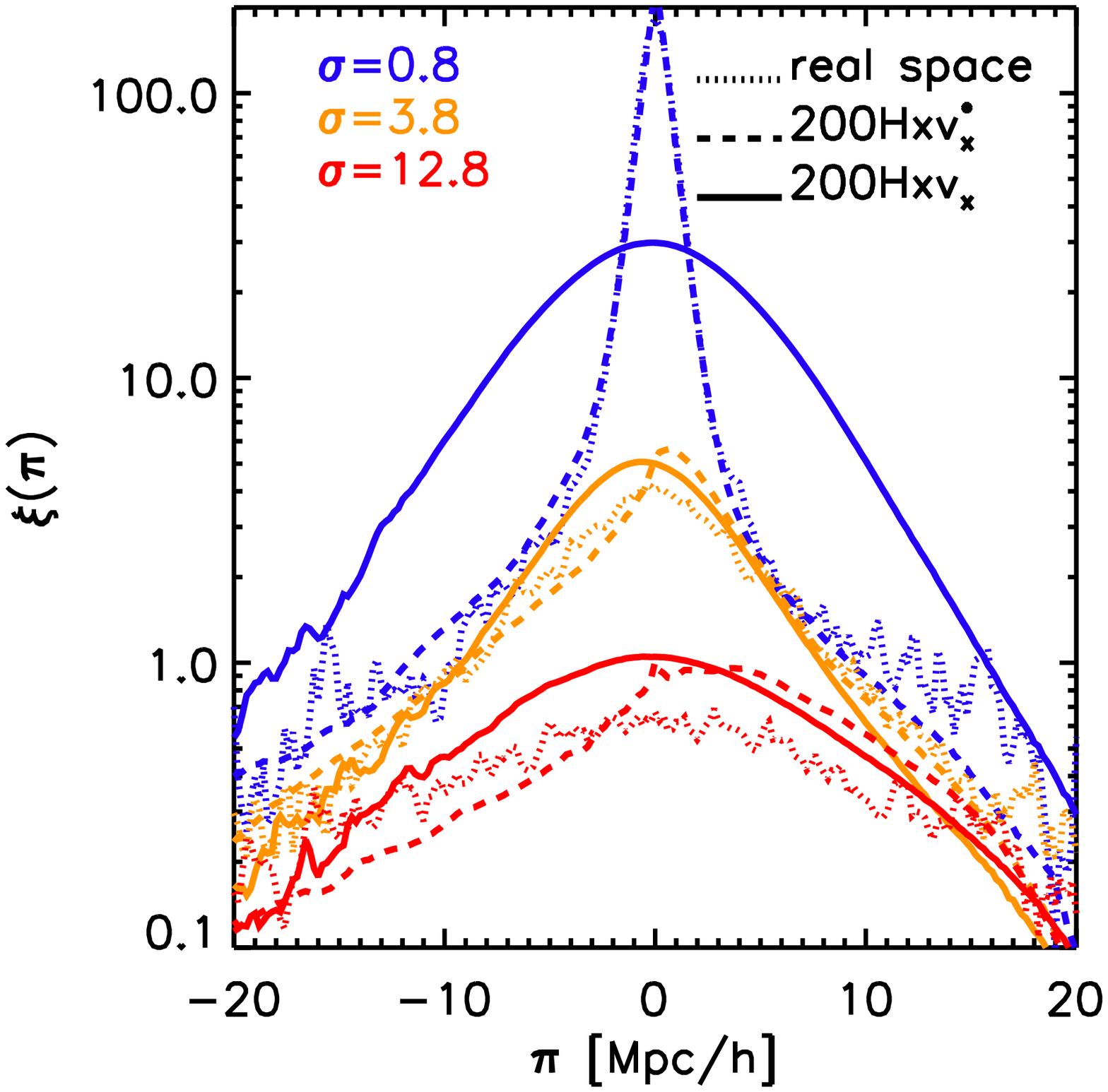}}
\caption{Left: averaged line-of-sight peculiar velocities of particles as a function of distance from the cluster centre in real space (dotted lines) and in velocity space (solid lines) at different projected distances indicated by the different colours. At small $\sigma$ values, the sign of the LOS  velocity in velocity space is flipped with respect to its real space version due to velocity dispersion and infall. Right: distribution of dark matter particles along the line of sight for three different cases. (1) real space (dotted lines); (2) with the line-of-sight real space positions of particles  perturbed by the second order term $200 Hxv_x$, where the boost factor of 200 is used for better illustration (dashed); (3) the line-of-sight velocity space positions of particles are perturbed by $200 Hxv_x$. The peaks of the particle distribution for case (1) are expected to be at the centre. In case (2), they are shifted to the positive $\pi$ direction (redshift) and the distribution is skewed due to the effect of the $200Hxv_x$ term. In case (3), at relatively small $\sigma$ values (the orange and blue curve), the peaks are shifted  to negative $\pi$ (blueshift) due to the fact that the sign of the LOS velocity is flipped in velocity space, as shown on the left-hand panel. These results are for haloes within the mass range of $M>10^{14}M_{\odot}/h$.}
\label{Fig:Hxdotx_sign}
\end{center}
\end{figure*}

Finally, in the right hand column of Fig.~\ref{FigRSD}, we have both the gravitational redshifts and peculiar velocities turned on. Both the correlation functions and the PDF's look essentially identical to the case of RSD only as the additional gravitational redshifts are much smaller than the peculiar velocity distortions. For the best-fit $\Phi^{\rm obs}$, even though they are noisy, we  see offsets of about 10~km/s when comparing the bottom-right panel with the bottom-middle panel. The difference between them, as shown by the solid curve in the bottom left panel, is {\it roughly} consistent with the dashed curve shown in the same panel, which is the GRedshift signal recovered in real space. 

In observations, the recovered gravitational redshift signal should be something like the bottom-right panel. 
It is affected by sample variance. The effect of sample variance can overwhelm the signal if the sample is too small. 
From simulations, we can effectively reduce sample variance by subtracting from the case of gravitational redshift + RSD the result of RSD only. This yields the GRedshift signal free from sample variance. We find this method is robust regardless of the amplitudes of the gravitational redshift signal. The result is shown by the solid curve in the bottom-left panel. 

To double check for the robustness of this method, we test using another technique to eliminate sample variance. 
We view each cluster from two opposite directions and stack them together before performing the fitting. 
This guarantees that each stacked cluster is perfectly symmetric along the line of sight in velocity space. 
The pure gravitational redshift signal can then be recovered. We find the recovered gravitational 
redshift signal from these two methods are consistent with each other. We will later apply them to measure the other quantities on the RHS of Eq.~(\ref{Eq:full2}). 
 
Quantitatively, the recovered GRedshift signal in velocity space (solid curve in the bottom-left panel of Fig.~\ref{FigRSD}) is found to be different to the real space version (dashed curve). This indicates the strong influence of the peculiar velocity on the 
observed GRedshift signal. Two more examples of this comparison are shown in Fig.~\ref{Fig:AllTerms}, where the pure gravitational redshift signals are shown by the red curves. The real space GRedshift signal recovered from the CGCF is consistent with the measurements shown in Fig.~\ref{Fig:Stack}. The troughs at $\sim 2$Mpc/$h$ indicate the 
impact of neighbours is again important. In velocity space however, those troughs no longer exist 
and the GRedshift profiles are very different from 
their real space counterparts. 

The difference of $\Phi^{\rm obs}$ in real and redshift space is not surprising. The observed redshift of particles or galaxies with large velocities relative to the cluster centre will appear far away from their original positions in the cluster system. This will 
alter their distributions along the line of sight, shifting the peaks of the PDFs relative to the cluster centre. The observed GRedshift signal 
in velocity space is therefore different from its original real space version. Note that due to the domination of the velocity dispersion over other effects of interest, all the other terms on the RHS of Eq.~(\ref{Eq:full2}) will also be significantly altered in velocity space.
In order to match observations, it is therefore important to make model predictions of this kind in velocity space.

\subsection{Other second order terms in the past the light cone}
\label{sec:2nd_Order_Term}

We now discuss in some detail the various terms occurring in Eq.~(\ref{Eq:full2}). The reader who is not interested in such details may skip to the conclusions.

Using the same technique as the previous subsection, we quantify the effect of the other terms on the RHS of 
Eq.~(\ref{Eq:full2}). In light of the strong impact of peculiar velocities on the predicted GRedshift signal shown previously, 
we show results only in velocity space for the other second order terms, but we have also checked explicitly their real space counterparts to gain a better understanding of the physics.

$\bullet$ $\Delta v^2$: Labelled as $\Delta v^2/2c$ in Fig.~\ref{Fig:AllTerms}, the special relativistic correction term always produces a redshift and so
is found to have 
the opposite sign to the GRedshift signal (blue curve), consistent with the results of \citep{Zhao2013}. However, the amplitude of the signal turns out somewhat smaller, i.e. at the sub-km/s level. 
When we examine its real space version, we find that there is a peak within the virial radius and its amplitude is approximately a factor of 2 larger than that in velocity space. 
This can be understood by the fact that particles having large peculiar velocities are displaced in $v$-space from their original locations. The $v$-space version therefore 
turns out to be smoother and have no obvious peak. 

$\bullet$ $[(\Delta v_x)^2-\Delta v_x^2]/2c$: This can also be written as $(v_{xc}^2-v_{xc}v_x)/c$ from which it can be seen that, by definition, it vanishes at the position of the BCG. At non-zero distance from the cluster centre, the second term should be very small when averaged over a large sample, leaving $v_{xc}^2$ as the dominant term. So this is effectively the special relativistic correction arising from the non-zero velocity dispersion of the BCG. From the orange curves in Fig. ~\ref{Fig:AllTerms}, we see it is nearly a constant as expected.

$\bullet$ $Hx v_x/c$: The term $Hxv_x$ is shown by the cyan curves. It is the product of the radial Hubble flow with the line-of-sight peculiar velocity. 
In the virialised region $Hx$ and $v_x$ are uncorrelated because the peculiar velocities are random. In the outskirts of a cluster, they are anti-correlated because of infall,i.e. $Hx$ is positive and $v_x$ is negative. This remains the same until the peculiar velocities drop to zero at very large distances, where they are back to no correlation. Initially, one might expect that in real space, this induces negative redshifts (blueshifts) 
with respect to the cluster centre, the same as the GRedshift signal. This is true for individual particles or galaxies, but what we find is that peak of the particle distribution is actually redshifted.
However, when switch to velocity space, the sign of this term 
is reversed again to become a blueshift. 

%the how the peaks of the particle distribution are shifted depends also on the 
%density profiles along the LOS, and there are more subtitles when we turn to $v$-space. 
Fig.~\ref{Fig:Hxdotx_sign}
shows an example to explain all these subtleties. Initially, the PDFs of particles along the LOS  are symmetric about the centre 
in real space, as shown by dotted curves in the right-hand panel. When adding the term  $200 Hxv_x$ (dashed curves), the PDFs are skewed and the 
peaks are shifted to the positive $\pi$ direction, even though individual particles move in the negative $\pi$ direction. 
This happens because of the joint effect of the amplitude of $Hxv_x$ increasing and the amplitude of the PDF decreasing with increasing LOS distance.  
At large positive $\pi$, particles are shifted towards the centre, which increases the amplitude of the PDF near the centre. At large negative $\pi$,  particles are shifted away from 
the centre, causing a decrease of the amplitude of the PDF. The consequence is that the peak of the PDF is shifted in the positive $\pi$ direction. The shift is more pronounced at large $\sigma$ (orange and red dashed curves) as the amplitudes of $200 Hxv_x$ is larger. 

In velocity space (solid curves), it is noticeable for the orange curve that the peak of the PDF is 
shifted in the negative $\pi$ direction. This is because the sign of the average velocity along the LOS $v_x$ is flipped in velocity space at relatively small $\sigma$ values. This is
shown by the left-hand panel of  Fig.~\ref{Fig:Hxdotx_sign},  the LOS $v_x$ is negative at all values of $\sigma$ as expected from the infall motion of mass towards the 
stacked cluster centre. However random particle velocities close to the
cluster centre displace particles with positive velocity to positive
$\pi$ coordinate in velocity space and vice versa. This
reverses the correlation between $v_x$ and $\pi$.
 This only occurs at relatively small $\sigma$ values, i.e. $\sigma \lesssim 9\Mpc$.  Therefore, the effect of the term $Hxv_x$ in velocity space is to
cause blueshifts at $\sigma \lesssim 9\Mpc$, which is the same as the GRedshift effect, but it gives rise to redshifts at $\sigma \gtrsim 9\Mpc$. This is shown by the cyan line in Fig.~\ref{Fig:AllTerms}.

$\bullet$ $-xg_x$: The green curves show the effects of the term $-xg_x$, minus the product of the line-of-sight displacement and
the line-of-sight acceleration.  We can understand it as arising from the change of the velocity of the galaxy with respect to the cluster centre during the interval of look-back time between the galaxy and cluster centre.
It is defined to be zero at the cluster centre. 
At a non-zero projected distance from the cluster centre, $x$ and $g_x$ tend to be anti-correlated (for over-dense systems), i.e. the acceleration will decrease (becoming less negative) with 
increasing distance from the cluster centre. With the negative sign, we expect each individual particle (or galaxy) to be  
redshifted (positive redshift) with respect to the cluster centre. Initially, one may expect that in real space, this term will have the opposite sign to that of the GRedshift effect. 
However, for the same reasons as those for the $Hxv_x$ term,  the peak of the CGCF is found to be shifted towards negative $\pi$ side (far side) of the centre in real space, 
and the sign of the recovered $-xg_x$ term flips again at  $\sigma < 9$ Mpc/$h$ in velocity space due to the infall velocities and dispersion. Therefore, the sign of the measured signal 
for the $-xg_x$ term is redshift at $\sigma < 9$ Mpc/$h$ and blueshift at larger projected distances, as shown by the green curves in Fig.~\ref{Fig:AllTerms}.

The quadratic (in $x$) term that comes from the the combination of the background gravitational redshift and Doppler effects is assumed to be removed by fitting the background density ramp to the line-of-sight galaxy distribution around the cluster centre, as reasoned in Section~2. We therefore do not include it in our figure.

Finally, the contribution of all these terms to the overall redshift signal are shown by the brown curve in Fig.~\ref{Fig:AllTerms}. They reduce the amplitude of the GRedshift signal (red-solid curve) by approximately 0.5~km/s and 1~km/s for the two halo samples presented in Fig.~\ref{Fig:AllTerms}. This is relatively minor (as some of them cancel with each other) compared to the other two systematics (the impact of neighbours and the combined effect of velocity space) identified earlier. 
With the effective volume of $3\times$[0.5(Gpc/$h$)]$^3$, the expected observed GRedshift signal is shown by the black-solid curves. All the systematics are overwhelmed by sample variance, which is reflected by the strong fluctuations of the curves. 

\section{Discussion and Conclusions}
We have explored how the modelling of the gravitational redshift signal from stacked clusters is 
affected by a variety of systematics. 

$\bullet$ 
Since the GRedshift signal is a component on the observed redshift, we start by presenting the expression Eq.~(\ref{Eq:full2})
for the observed redshift on the past light cone of an observer including relativistic corrections. It is relative to the centre of a cluster, and is expressed in terms of properties on surfaces of constant proper time. 
The effect of the second order terms in this expression on 
the cluster-galaxy cross-correlation function
are quantified using N-body simulations. We find that the the gravitational redshift term causes 
the strongest asymmetry of the CGCF. The recovered GRedshift signal is biased high by approximately
0.5-1~km/s depending on the minimum halo mass due to neglecting the other second order terms. This is relatively minor compared to the other two other 
systematics we have found. 

$\bullet$ The underlying gravitational potentials are usually deeper where there is a concentration of galaxies, 
which indicates a concentration of mass. The fact that observations of GRedshift are galaxy-number 
weighted causes the observed GRedshift signal to be biased low compared to models where volume weighting is assumed. This bias does not go away even if the stacked cluster is perfectly spherically symmetric. The non-spherical distribution of galaxies in individual clusters and the complex cosmic-web structures surrounding the cluster cause the bias to persist at nearly all scales of interest. This bias is stronger for lower mass clusters as the chance of having more massive neighbouring structures is higher. A pronounced bump at approximately
2~Mpc/$h$ from the cluster centre is expected for the observed GRedshift profile due to this bias. However, 
the bump tends to be flattened in velocity space. 

$\bullet$ Peculiar velocities of galaxies are the most dominant feature in the CGCF. The measurement of the GRedshift signal is in essence conducted in velocity space. It is strongly influenced by peculiar velocities 
since the observed galaxies are shifted from their original locations, e.g. galaxies at the bottom of the potential may appear far away from the cluster centre due to velocity-space distortions. This tends to flatten the bump of the GRedshift profile caused by the impact of neighbouring structures as mentioned in the previous bullet point. It also affects the predictions for all the other second order terms in Eq.~(\ref{Eq:full2}).

$\bullet$ We find that the CGCF along the line of sight  associated with the GRedshift signal is highly non-Gaussian. Therefore, extracting the signal by using a Gaussian function to fit for the peak positions of the CGCF as done in \citet{Wojtak2011,Sadeh2015} and \citet{Jimeno2015} may not be the optimal. There may be room for improvement in future analysis of this kind. The box-size of the simulation we use in this study is relatively small. The methods we have developed allow us to extract the relatively weak signal free from sampling variance. Simulations with larger box-size will be needed to study the noise properties.

\section*{Acknowledgments}
YC was supported by funding from an STFC Consolidated Grant, the European Research Council under grant number 670193 and the Durham Junior Research Fellowship.
YC acknowledges a grant with the RCUK reference ST/F001166/1. SMC acknowledges the support of the STFC [ST/L00075X/1] and ERC [GA 267291]
The simulations and part of data analysis for this paper were performed 
using the DiRAC Data Centric system at Durham University,
operated by the Institute for Computational Cosmology on behalf of the
STFC DiRAC HPC Facility (http://www.dirac.ac.uk). This equipment was funded by
BIS National E-infrastructure capital grant ST/K00042X/1, STFC capital
grant ST/H008519/1, and STFC DiRAC Operations grant ST/K003267/1 and
Durham University. DiRAC is part of the National E-Infrastructure.
Part of the analysis was done on the Geryon cluster at the Centre for Astro-Engineering UC, which
received recent funding from QUIMAL 130008 and Fondequip AIC-57. Access to the simulations used in 
this paper can be obtained from the authors.

\bibliography{Gredshift}

\bibliographystyle{mn2e}

\end{document}